# Curvature-guided topology and self-assembly in chiral nematics and liquid-crystal colloids

Ivan I. Smalyukh[1,2,3,4,*] and Mykola Tasinkevych[4,5,6†]

[1]*Department of Physics, University of Colorado Boulder, CO 80309, USA.*

[2]*Department of Electrical, Computer, and Energy Engineering and Materials Science and Engineering Program, University of Colorado, Boulder, CO 80309, USA.*

[3]*Renewable and Sustainable Energy Institute, National Renewable Energy Laboratory and University of Colorado, Boulder, CO 80309, USA.*

[4]*International Institute for Sustainability with Knotted Chiral Meta Matter $WPI-SKCM2$, Hiroshima University, Higashi Hiroshima, Hiroshima 739-8531, Japan.*

[5]*Centro de Física Teórica e Computacional, Faculdade de Ciências, Universidade de Lisboa, 1749-016 Lisboa, Portugal.*

[6]*Departamento de Física, Faculdade de Ciências, Universidade de Lisboa, P-1749-016 Lisboa, Portugal.*

In soft condensed matter, curvature does more than simply distort an ordered medium: it helps select defect structures, redistribute elastic stress, bias chirality, and guide self-assembly. This review examines how curved, multiply connected, and knotted boundaries in liquid-crystal colloids and confined nematics generate topological defects and localized solitonic textures, and how these structures mediate interactions between mesoscale building blocks. We introduce a unifying framework based on genus, Euler characteristic, anchoring, and chirality, and use it to discuss spherical, handlebody, and boundary-bearing colloids, together with droplets and polymer-dispersed nematics of nontrivial topology. Particular emphasis is placed on the interplay of geometry and topology in determining boojums, disclination loops, hedgehog charges, and linked and knotted defect structures. We then turn to chiral systems hosting skyrmions, torons, hopfions, and related localized textures, highlighting how chirality and confinement stabilize three-dimensional topological states. Finally, we discuss how these concepts translate into design principles for controlled self-assembly, templating, and functional composite materials. More broadly, we argue that liquid-crystal colloids and confined nematics provide experimentally accessible model systems in which curvature, topology, and chirality can be harnessed as programmable tools for designing organized soft matter.

∗ivan.smalyukh@colorado.edu
†mtasinkevych@ciencias.ulisboa.pt

## 1. Introduction: curvature as a design field in soft matter

In many areas of condensed-matter physics, curvature appears primarily as a geometric modification of an established material system: a bent quantum wire, a curved magnetic film, a superconducting nanoarchitecture, or a strained two-dimensional crystal. In such settings, geometry alters the spectra, transport properties, or collective modes of a pre-existing structural framework. Soft condensed matter offers a complementary perspective. Here, curvature does more than perturb the response of a structure; it participates directly in selecting the symmetry and organization of the ordered state itself, including its associated topological defects, textures, and pathways of self-assembly [1,2].

Liquid crystals (LCs) are exceptionally suited to this perspective. Combining fluidity with long-range orientational order, LCs are acutely sensitive to confinement, boundary conditions, and external fields. When geometric frustration is imposed, the material does not respond through brittle failure or conventional plastic rearrangement; instead, it reorganizes its order-parameter field. Because the liquid-crystalline order is energetically soft—with elastic constants typically in the piconewton range—the director field is highly susceptible to both geometric constraints and external driving. This susceptibility ensures that surface anchoring at colloidal interfaces can readily dictate the bulk texture, while also allowing for easy reconfiguration by modest external voltages (on the order of 1–4 V). Furthermore, these systems are experimentally transparent: optical microscopy, nonlinear imaging, and laser manipulation can be combined with numerical modeling to reconstruct three-dimensional textures directly in real space [1,2]. This ability to observe, with sub-micrometer resolution, how geometry reorganizes order is a major advantage of soft matter in the broader study of curvature effects.

A defining characteristic of soft matter is that curvature rarely acts in isolation; its physical consequences are inextricably linked with topology, anchoring, chirality, and elasticity. Curved interfaces and multiply connected boundaries dictate preferred local orientations of the director, yet the extension of these orientations into the bulk is governed by global topological constraints. These constraints are expressed through the defect structure. Depending on the surface topology and the degree of confinement, the system's response may manifest as boojums, disclination loops, hedgehog defects, or more localized topological solitons [1,2]. In chiral systems, the landscape is even richer: curvature and confinement compete with the intrinsic twist of the medium to stabilize structures with no direct analog in flat, topologically trivial settings, such as torons, skyrmions, hopfions, heliknotons and related knotted or linked textures [3,4].

Consequently, soft matter becomes more than a classical counterpart to hard condensed-matter systems; it serves as a primary setting for the direct study of how geometry organizes fields. While in many hard-matter systems, nontrivial topological textures must be inferred indirectly through transport or scattering, liquid-crystal colloids allow one to create, manipulate, and visualize such structures in three dimensions. This makes soft matter a model environment for addressing fundamental questions concerning field topology, geometric frustration, and defect-mediated organization [1].

These structures are not merely passive signatures of curvature; they also mediate interactions. A colloidal inclusion produces an elastic deformation that can attract, repel, or orient other inclusions. When the inclusion possesses nontrivial topology—such as a handlebody geometry, a knot-like shape, or a surface with boundary—the induced distortions and defects inherit that complexity, opening interaction channels absent in conventional isotropic suspensions [2,5–7]. Curvature and topology thus become design variables for self-assembly, providing a clear example of how geometry is translated into interaction symmetry and, ultimately, into organized structure.

Recent research on liquid-crystal colloids has made this program increasingly systematic. On one hand, the study of elastic multipoles and defect-mediated binding has established a language for understanding director distortions as effective interactions between mesoscale objects [2]. On the other hand, the introduction of progressively more complex inclusions—such as colloidal handlebodies [5,6], surfaces with boundaries and apex boojums [7], and linked or knot-shaped particles [1]—has tested how topology reshapes admissible field configurations. These studies demonstrate that curvature in soft matter extends far beyond the classical problem of defects on curved shells; it encompasses embedded inclusions with nontrivial topology and three-dimensional chiral textures stabilized by the interplay of geometry and frustration [8,9].

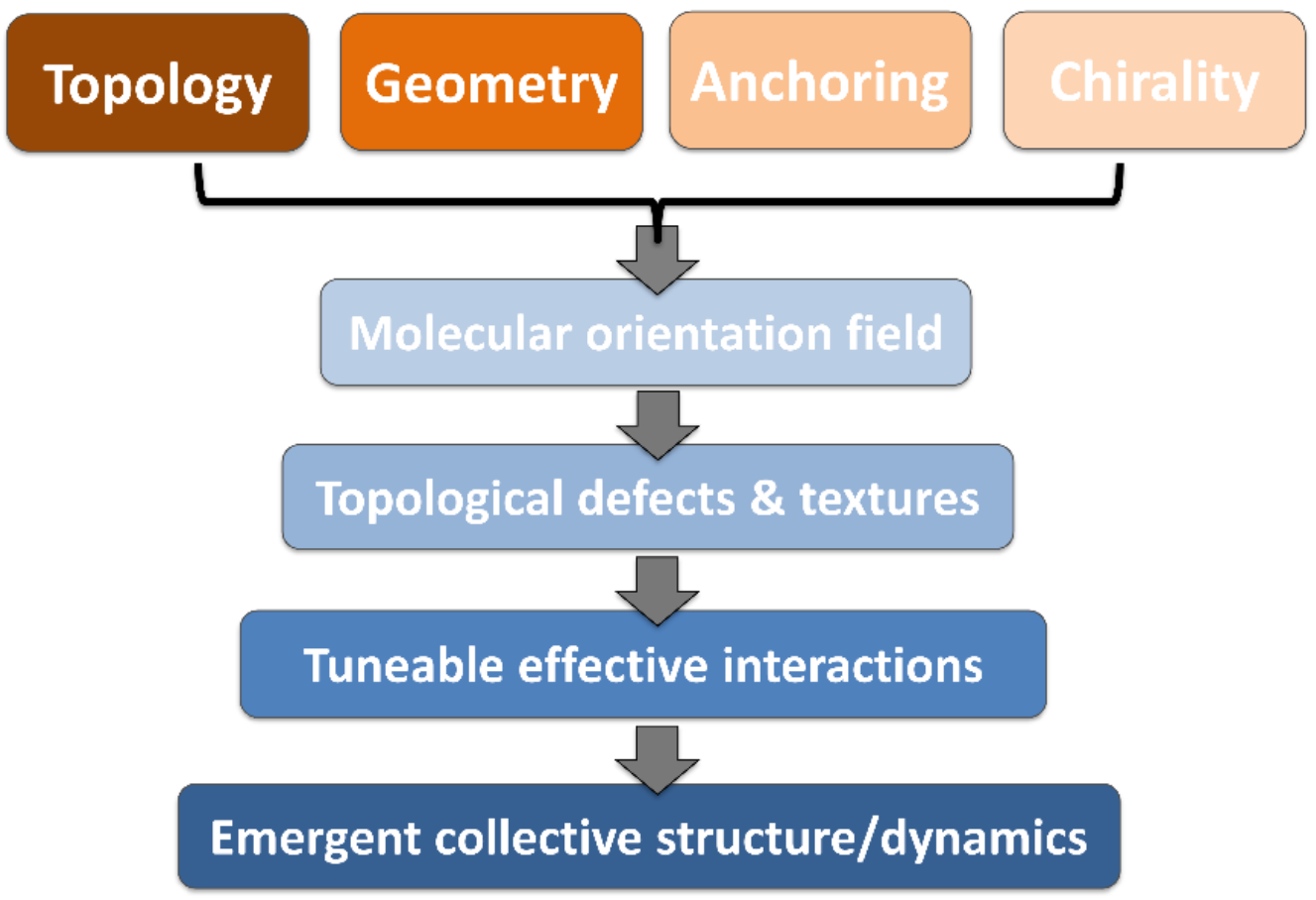

**Figure 1. Topology–geometry-anchoring-chirality design diagram.** Schematic overview showing how surface topology, geometry, and anchoring enriched by the material chirality combine to determine defect class, elastic interaction symmetry, and self-assembled outcomes in liquid-crystal colloids and confined nematics. Topology fixes global constraints through theorem-related topological invariants describing interplay of topologies of fields and surfaces; geometry and anchoring select defect realization; chirality enriches accessible textures.

This perspective shifts attention away from local bending alone. The relevant curvature effect is often the collective action of local geometry, global topology, anchoring, and elastic relaxation. These ingredients determine whether frustration is relieved by singular defects, continuous "escape," or the formation of texture-mediated mesoscale assemblies [6,8–10]. In soft matter, curvature is best understood not as an isolated geometric descriptor, but as a core component of a coupled design problem.

The approach adopted in this review is therefore broader than one centered only on free interfaces or nematic shells. Curvature is treated here as a design field acting through multiple channels: the curvature of confining boundaries, the geometry of embedded inclusions, the topology of surfaces with and without boundary, and the effective curvature of twisted textures in chiral media. We focus on how these environments generate and stabilize specific classes of structures—from disclinations around colloids to localized chiral solitons like torons and skyrmions [1,3,4,8,9]. We also emphasize the practical utility of this viewpoint: these structures encode interaction symmetry and provide experimentally accessible routes toward functional, adaptive composite materials.

In this context, the central question becomes: how can curvature be converted into organization? In liquid crystals, this progression is particularly transparent: curvature, topology, anchoring, and chirality determine the admissible textures; those textures define interactions; and those interactions generate functional mesostructures. This progression provides the organizing theme of this review, summarized schematically in figure 1. The aim is not to provide a fully exhaustive survey of all related work, but rather to explain and connect key concepts through representative examples, including several drawn from the authors' own research.

The phenomena reviewed below range from droplets whose topology forces defect lines through their bulk, to colloidal handlebodies and knots that imprint linked or knotted defects into surrounding nematics, to chiral solitons that self-assemble into moving crystalline states. Together, these examples illustrate a common principle: curvature, topology, anchoring, and chirality do not simply perturb orientational order; they can be used to design and organize it across scales.

## 2. Minimal design framework: curvature, topology, anchoring, chirality

Before turning to specific examples of droplets, handlebodies, knotted inclusions, and chiral localized textures, it is useful to establish the minimal framework used throughout this review. The purpose of this section is not to provide a comprehensive topological account of ordered media, but to introduce the concepts needed to connect geometry with defect structure and, ultimately, with self-assembly. In liquid crystals this connection is unusually direct: the geometry and topology of a confining surface or embedded inclusion impose orientational constraints, those constraints generate elastic frustration, and the system relieves that frustration through singular defects, escaped textures, or localized solitonic states [1,2,11–14]. The concepts introduced here are therefore not only classificatory; they provide the language needed to understand how curvature helps determine which ordered states can form in the systems reviewed below.

### 2.1. Geometry and topology of confining surfaces

We begin with the geometry of a surface embedded in three-dimensional space. Locally, such a surface is characterized by two principal curvatures, $k_1$ and $k_2$. From them one defines the mean curvature,

$$H = \frac{k_1 + k_2}{2},$$

and the Gaussian curvature,

$$K_G = k_1 k_2.$$

These local quantities matter because they influence how orientational order can be accommodated near the surface. Regions of positive, negative, or vanishing Gaussian curvature generally favor different director distortions and can bias the preferred positions of defects.

Local curvature, however, is only part of the picture. Equally important is the global topology of the surface. For a closed orientable surface of genus $g$, the Euler characteristic is

$$\chi = 2 - 2g. \tag{1}$$

Thus, a sphere has $\chi = 2$, a torus has $\chi = 0$, and a double torus has $\chi = -2$. For an orientable surface with genus $g$ and $b$ boundary components, the Euler characteristic is modified to

$$\chi = 2 - 2g - b. \tag{2}$$

Here $b$ is the number of distinct boundary loops: for example, $b = 0$ for a closed surface, $b = 1$ for a disk-like surface with a single edge, and $b = 2$ for an annulus.

This distinction is important throughout the discussion below. Closed colloidal handlebodies and closed droplets belong to the first class, whereas thin foil particles, pyramidal cones, and related colloids that realize surfaces with boundaries belong to the second [1,2,7]. These two classes illustrate a useful separation between geometry and topology. Local curvature acts by selecting regions where distortions or defects are energetically favoured, whereas the Euler characteristic imposes global constraints on the total topological charge that the director field must accommodate. Much of the physics reviewed below arises from the interplay between these two levels: geometry biases the placement and morphology of textures, while topology restricts the allowed defect budget.

### 2.2. Order parameter, elasticity, and anchoring

The systems considered here are mainly nematic liquid crystals, whose orientational order is described by a unit director field $\mathbf{n}(\mathbf{r})$, with the nonpolar identification $\mathbf{n} \equiv -\mathbf{n}$. This nonpolar nature is fundamental. It distinguishes nematics from ordinary vector fields and permits the existence of half-integer disclinations. Topologically, the relevant order-parameter space is therefore not the sphere $S^2$, but the real projective plane $RP^2 \cong S^2/\mathbb{Z}_2$ [1,11–13].

At the continuum level, distortions of the director field are described by the Frank-Oseen free energy,

$$F_{\mathrm{el}} = \int_V \left[\frac{K_{11}}{2}(\nabla \cdot \mathbf{n})^2 + \frac{K_{22}}{2}(\mathbf{n} \cdot \nabla \times \mathbf{n})^2 + \frac{K_{33}}{2}\,|\,\mathbf{n} \times (\nabla \times \mathbf{n})\,|^2\right] dV, \tag{3}$$

where $K_{11}$, $K_{22}$, and $K_{33}$ are the splay, twist, and bend elastic constants, respectively [2,11]. This expression is sufficient for most of the discussion below, although in some situations additional contributions, such as the saddle-splay term, may play an important role.

The director field is further constrained by surface anchoring. For homeotropic anchoring, with preferred orientation along the outward unit normal $\boldsymbol{\nu}$, the simplest Rapini-Papoular form is

$$F_{\mathrm{s}} = \frac{W}{2}\int_S [1 - (\mathbf{n} \cdot \boldsymbol{\nu})^2]\, dS, \tag{4}$$

where $W$ is the anchoring coefficient [2,11]. Similar forms can be written for tangential or conical anchoring by replacing the preferred surface orientation.

The balance between elastic and anchoring effects is often characterized by the surface extrapolation length

$$\xi_e = \frac{K}{W}, \tag{5}$$

where $K$ is a representative elastic constant. This length provides a simple measure of whether a particle or surface feature is large enough to impose its preferred boundary condition effectively. If its characteristic size is much smaller than $\xi_e$, the director may deviate from the preferred anchoring without paying a large energetic penalty. If it is much larger than $\xi_e$, the anchoring is effectively strong, and the system instead pays elastic and defect energies to satisfy the boundary condition [2]. This is one reason why micron-sized inclusions and confined droplets are so effective at revealing topological behavior: their geometry is strongly transmitted to the orientational field.

Three anchoring classes recur throughout this review. Homeotropic anchoring favours alignment of $\mathbf{n}$ along the surface normal; tangential anchoring confines $\mathbf{n}$ to the local tangent plane; and conical or tilted anchoring selects a fixed oblique alignment angle relative to the surface normal. Because anchoring determines how surface geometry is transmitted into the bulk director field, the same particle topology can generate very different defect structures under different interfacial conditions.

### 2.3. Surface defects and bulk topological charge

The most practical topological relations in these systems are the ones that connect the topology of a surface to the defect content of the surrounding orientational field. These relations do not determine a unique director configuration, but they sharply restrict the class of admissible states. Once that topological requirement is fixed, geometry, anchoring, and elasticity determine how it is realized in practice.

The simplest such relation concerns the projection of the director onto a confining surface. If $\mathbf{n}_s(\mathbf{r})$ denotes the tangential projection of the director and $s_i$are the winding numbers of the corresponding surface point defects, then the Poincaré-Hopf theorem gives

$$\sum_i s_i = \chi. \tag{6}$$

This relation provides the basic rule for boojums on colloids and droplets. For a sphere with tangential anchoring, $\chi = 2$, so the projected field at the particle's surface must carry net surface charge $+2$, commonly realized as two $s = +1$ boojums. For a torus, $\chi = 0$, so no net surface defect charge is required; the surface may be defect-free or may contain self-compensating positive- and negative-charge defects [6].

A second topological constraint concerns the bulk defects induced by a closed inclusion or confinement surface. If the director field at that surface is globally vectorized, the resulting mapping to $S^2$ assigns an effective hedgehog charge

$$m_c = \pm\frac{\chi}{2}, \tag{7}$$

where the sign depends on the arbitrary but globally consistent choice of vectorization of the nonpolar director [5]. The bulk defects induced in the surrounding nematic must then compensate this charge:

$$\sum_i m_i = -m_c. \tag{8}$$

For a sphere, $m_c = \pm 1$, so the surrounding bulk must carry the opposite unit charge, realized for example by a hyperbolic hedgehog or by a disclination loop with the same effective charge. For higher-genus inclusions or droplets, Eqs. (7) and (8) show that the total compensating charge is fixed by topology, even though the actual realization of that charge may vary.

To make this explicit, the degree of a vectorized director field on a closed surface $S$ can be written as

$$m = \frac{1}{4\pi}\int_S \mathbf{n}\cdot(\partial_1\mathbf{n}\times\partial_2\mathbf{n})\, dx_1 dx_2, \tag{9}$$

where $x_1$ and $x_2$ parametrize the surface [12]. In practice this quantity is often referred to as the hedgehog charge. Although it is not usually computed directly in experiment, it underlies the charge assignments used when discussing colloidal handlebodies and other topological inclusions.

Two points should be emphasized. First, topology fixes only the net defect content; it does not uniquely specify the morphology by which that content is realized. The required charge may appear in the form of point defects, loops, several defects whose charges sum appropriately, or partly escaped textures. Second, in a nematic the nonpolar nature of the order parameter makes half-integer disclinations possible. This greatly enlarges the set of admissible structures and is one reason linked and knotted defects arise so naturally in these systems [1,9].

### 2.4. Homotopy viewpoint

Although the discussion below remains mainly geometrical and physical in emphasis, a few standard results from homotopy theory are useful for orientation. For a three-dimensional polar vector field with target space $S^2$, one has $\pi_2(S^2) = \mathbb{Z}$, so point defects carry integer topological charge, whereas $\pi_1(S^2) = 0$, so isolated stable line defects are not topologically protected. In nematics, however, the order-parameter space is $RP^2$, and

$$\pi_1(RP^2) = \mathbb{Z}_2, \pi_2(RP^2) = \mathbb{Z}. \tag{10}$$

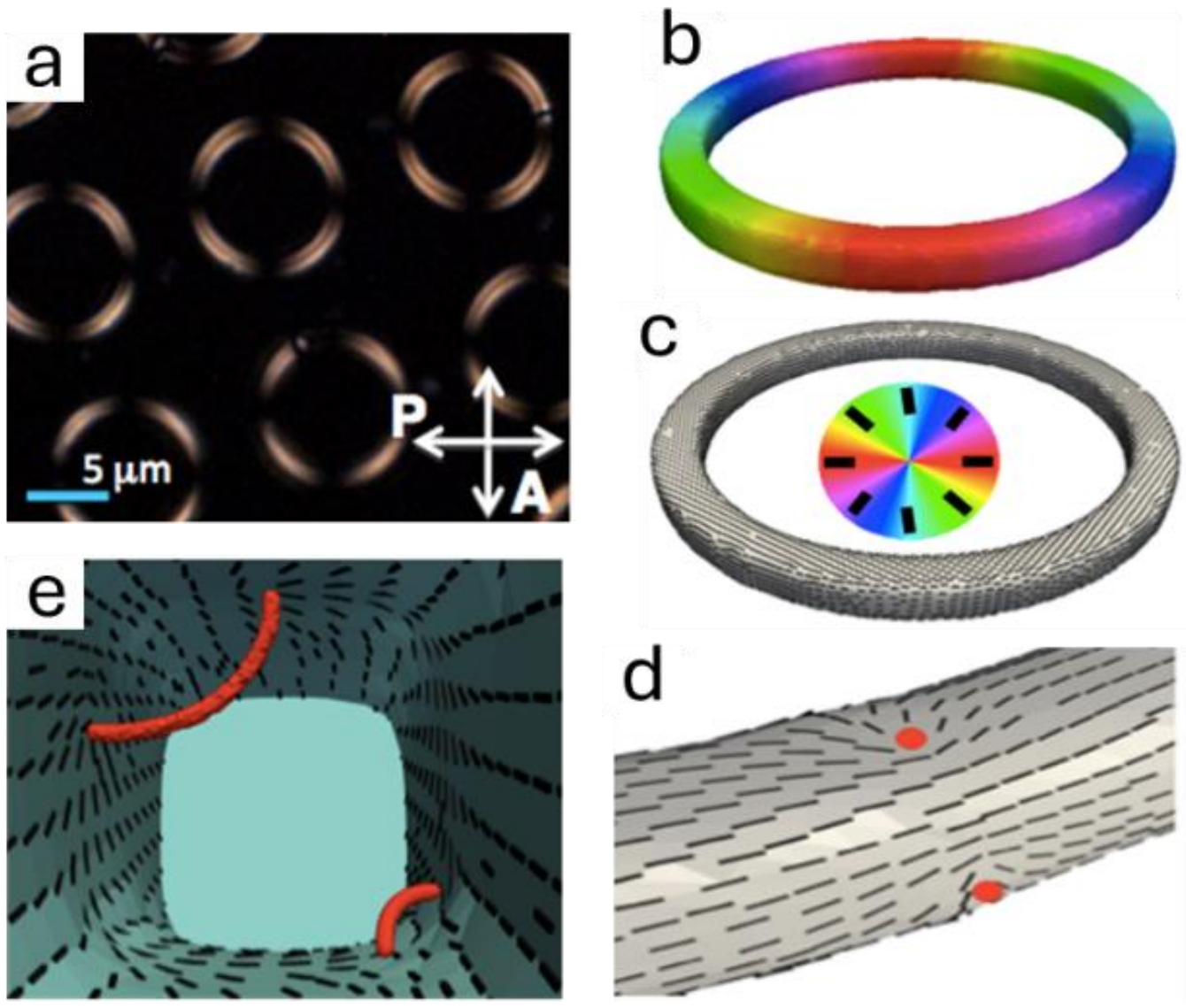

**Figure 2. Nematic LC droplets confined by surfaces of genus $g = 1$ with tangential anchoring**. Topological polymer dispersed LCs with genus $g = 1$ nematic drops. a, A polarising optical microscopy micrograph obtained between crossed polarizers shown using white double arrows. b, three-dimensional construction depicting director field $\mathbf{n}(\mathbf{r})$ at the surface of a torus droplet based on the results of numerical modeling, with the color-coded scheme of azimuthal orientations shown in the inset of c. c, Computer-simulated $\mathbf{n}(\mathbf{r})$ of a defect-free ground-state configuration depicted by rods on the surface of the droplet. d and e, Director structure in the vicinity of self-compensating topological defects found in metastable structural states of some droplets as depicted at the droplet surface interface, d, and in the interior of the droplet, e, where half-integer disclination lines, which start and end at certain points on the inner surface of the droplet, are shown by red tubes and filled circles representing surfaces of reduced scalar order parameter. Reproduced with permission from [8].

This means that both singular line defects and integer-charged point defects can exist in three-dimensional nematics [1,12,13]. Many of the structures encountered later—disclination loops around handlebody inclusions, linked defects in multiply connected droplets, and knotted singular lines—follow directly from this fact. We note also that the simultaneous presence of point and line defects in nematics requires special attention as moving a point defect around a defect line can flip its sign [15], though such cases are outside the scope of this review.

For nonsingular textures, compactification of physical space gives further classifications. In two spatial dimensions, skyrmion-like structures are associated with mappings $S^2 \to S^2$, corresponding to $\pi_2(S^2)$, whereas fully three-dimensional hopfion-like structures correspond to $\pi_3(S^2) = \mathbb{Z}$. In ordinary nematics this classification must be handled with some care because of the identification $\mathbf{n} \equiv -\mathbf{n}$, but after suitable vectorization—or in vectorial liquid crystal ferromagnets and ferroelectrics —it becomes directly helpful [1,4].

The essential point for the present review is that confinement and chirality can stabilize not only singular defects, but also topological textures that are partly or fully nonsingular.

### 2.5. Chirality and frustration

Chirality introduces another control parameter because the locally preferred state is no longer uniform alignment, but twist. In a cholesteric liquid crystal, this tendency is characterized by the helical wave number $q_0 = 2\pi/p_0$, where $p_0$ is the equilibrium pitch. The corresponding twist contribution to the free energy can be written as

$$F_{\rm ch} = \int_V \frac{K_{22}}{2} (\mathbf{n} \cdot \nabla \times \mathbf{n} + q_0)^2 dV, \tag{11}$$

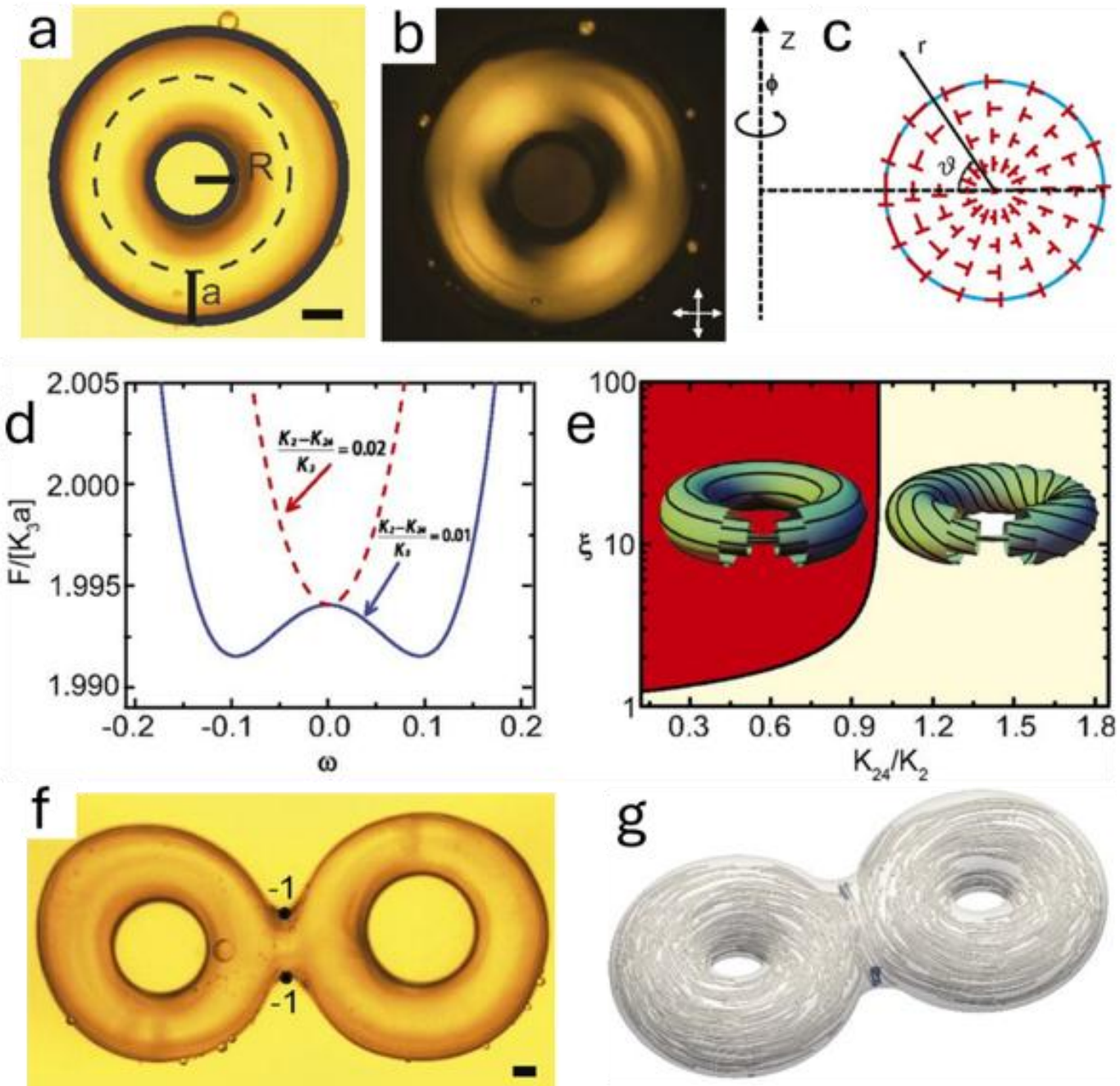

**Figure 3. Multiply connected nematic droplets: tangential anchoring**. A bright field, a, and cross-polarizers, b, views of a toroidal droplet of nematic liquid crystal, with the tube and the inner radii $a$ and $R$, respectively; scale bar is $100\mu m$. c, A cross-section view of a doubly twisted nematic configuration inside the torus, where nails represent the out-of-plane tilt of the director. The full three-dimensional director field is axially symmetric with respect to the $Z$ axis. d, Frank-Oseen elastic free energy as a function of the variational parameter $\omega$, quantifying a degree of the director twist (see [16] for details), with the torus aspect ratio $\xi = (R + a)/a$ =5, and two different combinations of the twist, $K_2$, bend, $K_3$, and saddle-splay $K_{24}$, elastic constants. The red dashed curve corresponds to untwisted ($\omega = 0$ ) axial director configurations represented by the red region in the $(K_{24}/K_2, \xi)$ configurational diagram in panel e. While the blue solid curve in d has two symmetric minima at $\omega \approx \pm 0.1$, which correspond to the two possible handednesses of the doubly twisted configurations represented by the yellow region in the diagram e. f, Top bright field view of a double torus. Solid dark circles represent two $-1$ surface defects; scale bar is $100\mu m$. g, A numerical director configuration confined to a double torus, with two $-1$ surface defects, located at the outermost regions of negative Gaussian curvature. Reproduced with permission from [16].

so that the free-energy minimum favors a twisted state rather than uniform alignment [11]. Once chirality is present, confinement becomes a direct source of frustration. A cell with homeotropic or planar boundary conditions may favor a nearly uniform director at its surfaces, while the bulk favors twist with wave number $q_0$. The resulting competition, which maps onto the competition between linear and nonlinear effects in the branch of nonlinear physics studying solitons, can stabilize localized twisted structures embedded in an unwound background. Among the experimentally important examples are skyrmions, torons, and hopfions, as well as defect loops and knotted director streamlines [1,3,4]. In the examples below, the ratio $d/p_0$, where $d$ is the confinement thickness, is one of the main dimensionless control parameters, alongside anchoring strength and elastic anisotropy. More broadly, chirality does more than energetically stabilize additional topological structures: it changes which localized textures can be selected, stabilized, and reconfigured under confinement.

### 2.6. Working framework for the rest of the review

The ideas introduced above can be summarized in a simple chain of logic progression. Geometry fixes local curvature and boundary orientation. Topology constrains the global defect budget through $\chi$, genus, and the number of boundary components. Anchoring translates surface geometry into preferred director alignment. Elasticity determines the cost of accommodating that alignment in the bulk. Chirality adds a competing preference for twist. The structures actually observed—boojums, disclinations, hedgehogs, escaped configurations, skyrmions or torons —are the outcome of this combined topological and energetic selection process.

This framework is used repeatedly in the sections below. In multiply connected droplets, Eqs. (6)–(9) determine the possible defect content, while geometry and elasticity select among several realizations. For colloidal inclusions with nontrivial topology, the same relations explain why handlebodies, knot-shaped particles, and surfaces with boundary induce qualitatively different director fields from those around ordinary spheres. In chiral confined systems, Eq. (11) explains how frustration between twist and boundary conditions can stabilize localized topological textures. In all of these cases, the essential point remains the same: curvature does not act alone, but rather acts as part of a coupled topological-elastic-chiral mechanism that helps determine which ordered states can form.

## 3. Curved confinement in droplets and handlebodies

Curved confinement offers one of the clearest settings in which to observe how geometry and topology shape orientational order in soft matter. Unlike an embedded solid inclusion, a droplet defines the ordered domain itself: the confining surface is also the phase boundary, and its geometry therefore constrains the director field throughout the interior. For this reason, droplets and related polymer-stabilized structures are notably revealing experimental realizations of curvature-driven organization. In the simplest spherical case, the relation between topology and defect content is already familiar from the classical literature on nematic droplets and shells. Here, however, the emphasis is on the less trivial case of non-simply connected confinement, such as toroidal and higher-genus domains, where curvature, topology, anchoring, and elastic anisotropy together determine whether frustration is relieved through surface boojums, bulk disclination lines, point defects, spontaneous twist, or nonsingular escaped configurations [1,8,9,16].

A useful way to organize the discussion is to distinguish between tangential and homeotropic anchoring at the confining surface. While the same global topological invariants apply to both cases, they govern different manifestations of the order-parameter field. Under tangential anchoring, the director is constrained to the surface plane, and the resulting topological constraints are expressed through the surface-bound defects (boojums) and their respective winding numbers. Under homeotropic anchoring, the surface acts as a source of net topological charge; this must be compensated within the bulk by point defects, disclination loops, or escaped textures to satisfy the total charge requirements of the system. In either case, a consistent principle emerges: global topology dictates the total topological charge that must be accommodated, while local geometry and free-energy minimization determine the specific configuration and the spatial placement of the defects used to achieve it.

### 3.1. Tangential anchoring: boojums, twist, and geometric selection

For tangential anchoring, the governing topological relation is Eq. (6): the winding numbers of the surface defects in the projected director field must sum to the Euler characteristic of the confining surface. A spherical LC droplet, with $\chi = 2$, must therefore carry total surface charge $+2$, while a toroidal droplet, with $\chi = 0$, is not required to host any net surface defects at all. This already marks an important departure from the spherical case. A toroidal droplet may be defect-free (figures 2a-c, and figures 3a and b), but it may equally well host self-compensating positive and negative boojums (figures 3f and g), provided the total charge remains consistent with the same topological rule [1,8].

The example of the torus is particularly instructive because it shows that the absence of a net topological requirement does not imply a trivial state. Experiments and simulations on toroidal nematic droplets with planar anchoring revealed that the director field may spontaneously acquire a twisted configuration, even when the liquid crystal itself is achiral (figure 3c) [16]. The origin of this behavior lies in the coupling between toroidal geometry and elastic anisotropy, notably the saddle-splay contribution (figure 3d). Rather than settling into a purely untwisted tangential texture, the system can lower its free energy by adopting a chiral, double-twist-like arrangement (figure 3e). In that sense, curved confinement does more than host defects: it can favor one handed organization over another and drive spontaneous mirror-symmetry breaking through elastic relaxation.

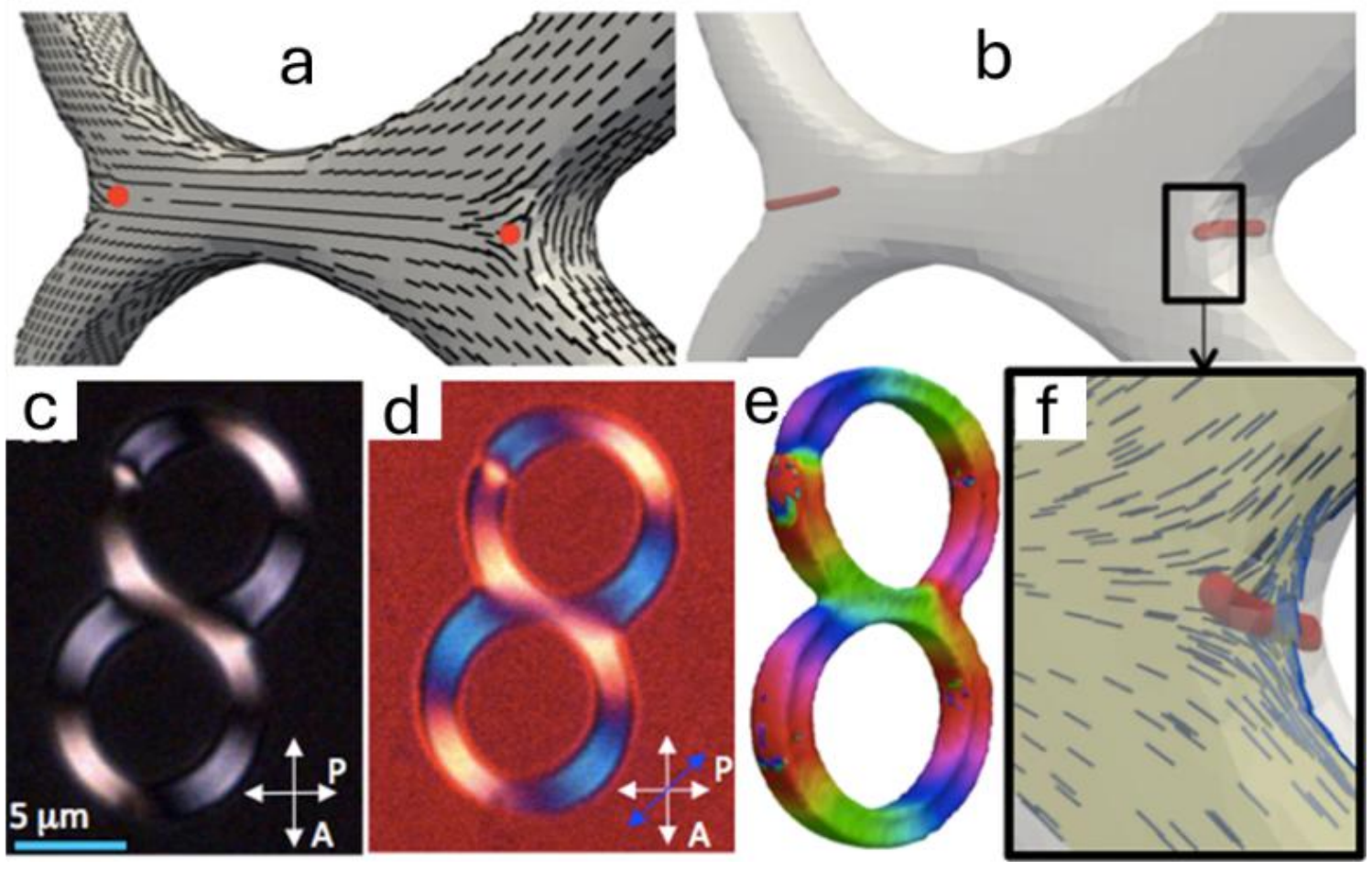

**Figure 4. Nematic droplets confined by surfaces of genus $g = 2$: Tangential anchoring**. a and b, Nematic field configurations and defects at the junction of two tori, with $\mathbf{n}(\mathbf{r})$ at the droplet surface shown by rods (a), and the line defects in the droplet bulk shown by red isosurfaces of reduced scalar order parameter. c and d, A metastable configuration with a pair of additional self-compensating defects visualized with polarizing optical microscopy between crossed polarizers without (c) and with (d) an additional phase retardation plate. e, Experimentally constructed 3D representation of $\mathbf{n}(\mathbf{r})$ at the surface of the $g = 2$ droplet with the two additional defects and corresponding to c and d. f, A detailed director configuration depicted in the vicinity of a defect in a midplane of the handlebody droplet. Reproduced with permission from [8].

Once the confinement acquires more than one handle, the topological constraint becomes correspondingly stronger. For a closed orientable surface of genus $g$, one has $\chi = 2 - 2g$, so each additional handle reduces the required net surface charge by two units. A genus-two droplet must therefore carry total surface winding number $-2$, and higher-genus droplets must accommodate progressively more negative total charge. In practice, that charge is not distributed uniformly. Surface defects tend to localize in regions where their elastic cost is reduced, such as saddle-like areas or geometrically selected junctions between handles (figures 3f and 3g) [16]. Topology, however, fixes only the total defect content; it does not prescribe how this requirement must be realized. In nematics, the nonpolar symmetry of the director further enlarges the set of admissible configurations by allowing half-integer disclinations. In the micronsize droplets shown in figure 4, the topologically required defects localize in the junction region between the two handles, similarly to the case of millimetre size droplets shown in figures 3f and g. Strikingly, unlike larger handlebody droplets where boojums are often observed, these microdroplets are dominated by half-integer singular lines spanning the droplet thickness rather than by surface point defects (figures 2d and e; figures 4b and f). The most common configuration contains two $s = -1/2$ defect lines in the inter-handle junction, which together satisfy the required net winding. In addition to these topology-required lines, one occasionally finds extra self-compensating defect pairs, corresponding to metastable states also reproduced by numerical modelling (figures 4c-e). Their persistence is naturally understood from the energetic cost of extending and annihilating defect lines: although such pairs are not required by topology, the droplet geometry can stabilize them by making annihilation pathways unfavorable. Thus, for $g = 2$ confinement, the observed textures illustrate particularly clearly the distinction between the global topological requirement, fixed by $\chi$, and the specific defect morphology, selected by geometry and free-energy minimization.

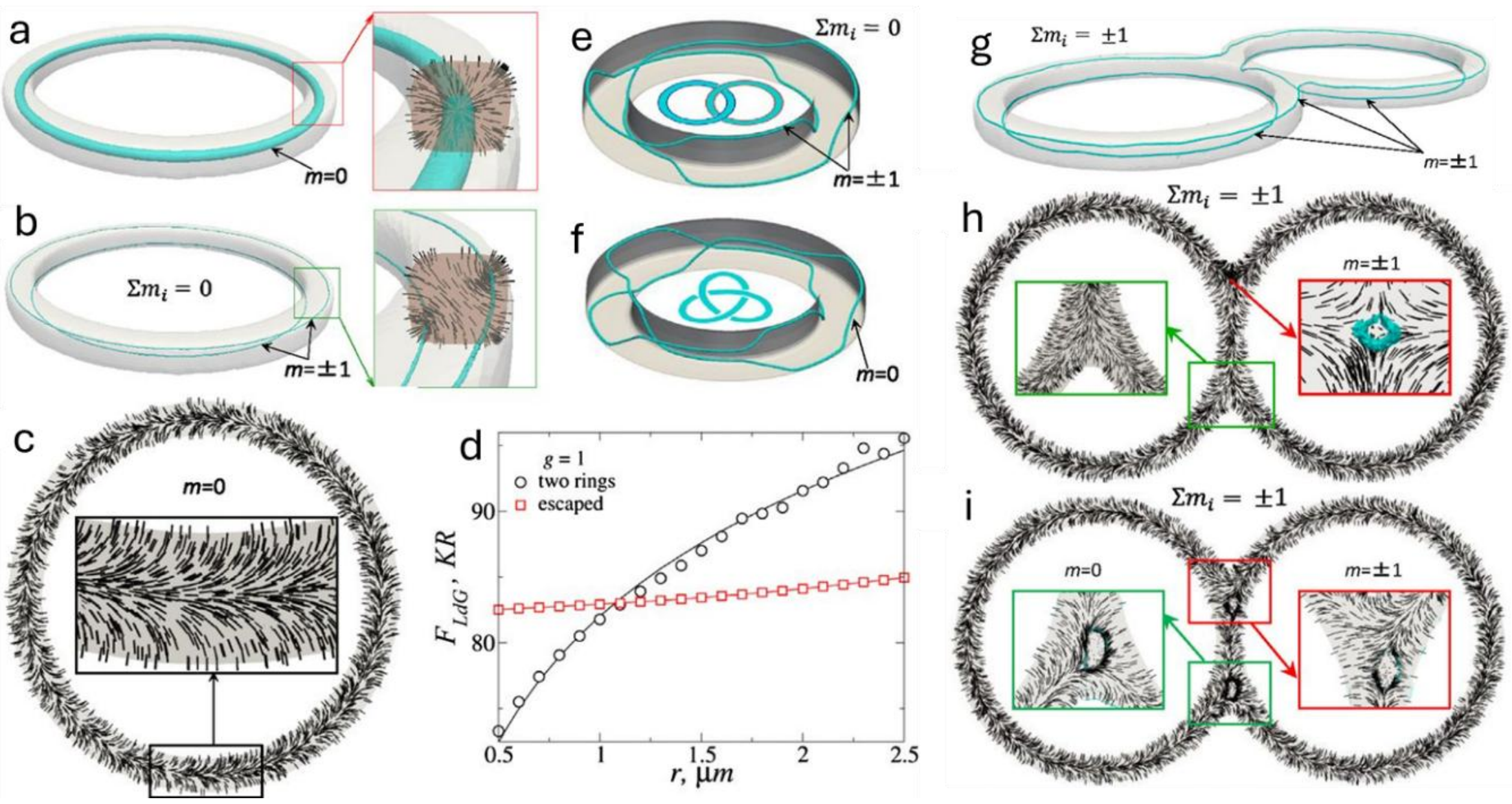

**Figure 5. Nematic LC confined by multiply connected surfaces**: **homeotropic anchoring**. LC configurations for confining surfaces at $g = 1$, exhibiting a single s = 1 (a) and two s = 1/2 disclination rings (b), shown as blue isosurfaces of a certain (reduced) scalar order parameter. Insets show cross-section views of $\mathbf{n}(\mathbf{r})$ around the disclination lines. e, A Hopf link and, f, trefoil T(3; 2) torus knot of half-integer disclination loops depicted by blue isosurfaces of a reduced scalar order parameter. c, $\mathbf{n}(\mathbf{r})$ in a plane of a torus with an escaped structure. d, Landau–de Gennes free energy as a function of the torus tube radius $a$ and for the two competing director configurations – the double rings, and the escaped one. g, $g = 2$ confinement resulting in a configuration with two half-integer disclination lines. h and i, depict for $g = 2$ escaped LC director structures in the midplanes of the double-tori. Insets focus on the junction regions and blue tubes correspond to reduced scalar order parameter isosurfaces. Hyperbolic hedgehog defect with a unit topological charge has a split ring-like core structure (red-framed inset in h). Also shown are, disclination loops with zero (green-framed) and unit (red-framed) topological charges. All the configurations are obtained via numerical minimization of the Landau-de Gennes free energy [9]. Reproduced with permission from [9].

This is precisely why multiply connected droplets are so revealing. They make it possible to disentangle local geometric preference from global topological constraint in an experimentally accessible way. The Euler characteristic determines what total charge must be present, but the actual arrangement of boojums reflects curvature variations, elastic anisotropy, and the possibility of twist or split-core structures. Such droplets are therefore more than striking visual realizations of topological rules; they are good testbeds for understanding how curvature landscapes bias defect placement in orientationally ordered media more generally.

### 3.2. Homeotropic anchoring: bulk compensation, linking, and escape

With homeotropic anchoring, the focus shifts from surface-bound defects to the bulk singularities or escaped textures required for the topological compensation of the charge induced by the confining surface. As discussed in Section 2, the vectorized director field at a closed orientable boundary carries the effective hedgehog charge given by Eq. (7), and the bulk must compensate it according to Eq. (8). For a surface of genus $g$, the magnitude of the required compensating bulk charge is therefore $|1 - g|$ [1,5,8,9]. This immediately clarifies the toroidal case. Since a torus has $\chi = 0$, the net compensating bulk charge must vanish. Yet, as in the tangential case, vanishing total charge does not imply a trivial state. The droplet may realize this condition through no singular defects at all, through self-compensating combinations of point defects and disclination loops, or through nonsingular escaped textures (figure 5). This multiplicity of admissible states is one of the characteristic features of higher-genus confinement in nematics. In studies of homeotropic nematic drops with handles, the same toroidal topology was shown to support singular loops (figures 5a,b, and g), linked defect structures (figure 5e), trefoil-like knotted configurations (figure 5f), and nonsingular escaped states (figures 5c, h, and i), depending on droplet size, aspect ratio, and elastic parameters [9].

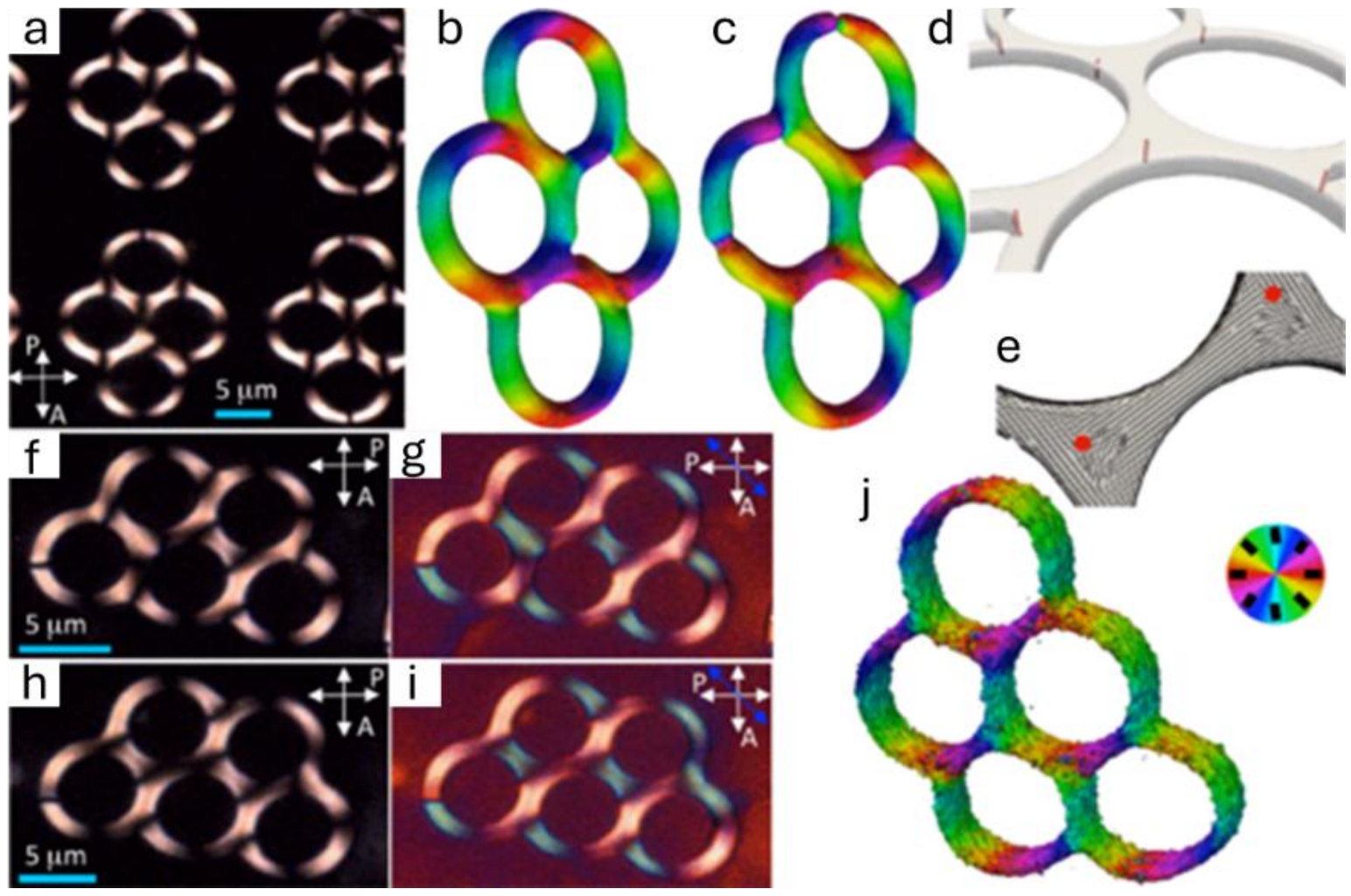

**Figure 6. Topological polymer dispersed LCs with genus $g = 4, 5$ nematic drops**. a, Polarizing optical microscopy (POM) micrograph, b and c, three-dimensional reconstructions of the surface director field for $g = 4$ droplets; the director orientations are color-coded according to the legend in j. d and e show the corresponding results of numerical minimization of Landau-de-Gennes free energy, highlighting half-integer defect lines in the regions of junction of different tori, d, and details of the director field at the LC-polymer interface, e, shown for one of such regions with the surface-pinned defect lines marked by the red circles. f-i show POM micrographs of two $g = 5$ drops taken between crossed polarizers without (f, h) and with (g, i) an additional phase retardation plate. (j) reconstruction of $\mathbf{n}(\mathbf{r})$ at the surface of a $g = 5$ drop. Reproduced with permission from [8].

For higher genus, the net topological charge dictated by the manifold becomes non-zero, and the competition between different defect configurations becomes correspondingly richer. Small or strongly frustrated droplets typically satisfy the required topological charge through singular point defects or closed disclination loops. In larger droplets, however, the system can often lower its free energy through partial or complete escape of the director field into the third dimension, leaving the remaining topological charge localized near handle junctions or other geometrically selected regions [9]. The essential point is that topology does not prescribe a unique morphology; instead, it imposes a global constraint that may be satisfied by multiple local minima within the free-energy landscape.

One of the most striking features of homeotropic handlebody droplets is that their singular structures can go far beyond isolated point defects or simple defect rings. Because the nematic director is nonpolar, half-integer disclinations are topologically acceptable, and in multiply connected droplets these can close into loops, link with one another, or even knot [1,9]. Such structures would not arise in the same way in a polar vector field, where isolated half-integer lines are forbidden. In nematics, by contrast, the combination of nonpolarity and three-dimensional curved confinement makes linked and knotted bulk defects a natural outcome of topological frustration.

### 3.3. Topological polymer-dispersed liquid crystals as geometric playground

A remarkably versatile realization of these ideas is provided by topological polymer-dispersed liquid crystals (TPDLCs), in which a polymer matrix stabilizes nematic domains of nontrivial topology and, for the case of tangential anchoring, pins defect lines to handlebody confining surfaces [8], figures 2 and 6. These systems are important because they decouple topological design from the surface-tension constraints of free droplets. Rather than relying on a liquid interface to stabilize a torus or multi-handle shape, one can impose the desired topology directly through a polymer scaffold and then examine the resulting director textures under much greater geometric control.

The key result is that the confining geometry does not simply trap defects that would otherwise exist. It actively selects and organizes them. For tangential boundary conditions, the surface defect content remains governed by Eq. (6), but the actual defect positions are pinned to geometrically favorable regions such as handle junctions or saddle-like surface patches [8]. For homeotropic boundary conditions, the same confining topological surface structures may instead host bulk disclination lines, loops, or escaped textures, depending on the elastic and geometric parameters [9], figure 5. The polymer matrix, therefore, acts as a topological template, converting connectivity and curvature into repeatable three-dimensional director morphologies.

This templating viewpoint is important for the broader argument of this review. It shows that curved confinement is not only a way of revealing topological rules in a soft medium, but also a practical method for writing topological states into matter. Because these states are optically visible and, in many cases, reconfigurable by external fields or thermal cycling, polymer-stabilized handlebody domains connect abstract topological constraints to experimentally addressable soft-matter architectures.

### 3.4. Curved confinement as a design principle

Taken together, the examples above point to a broader interpretation of the interplay between the topologies of fields and surfaces at higher-genus confinement. A torus, double torus, or more complex handlebody is physically significant not only for its topological distinction from a sphere, but because it fundamentally reshapes the free-energy landscape of orientational order. The coexistence of positive and negative Gaussian curvature, the presence of handle junctions, and the nature of nontrivial connectivity create preferred regions for defect localization, twist penetration, and director escape. At the same time, global topology dictates net topological invariants through Eqs. (6)–(8). The resulting textures are therefore neither arbitrary nor uniquely fixed; rather, they belong to a restricted set of topologically admissible configurations, from which elasticity, anchoring, and local geometry select the realized state.

Curved confinement should, therefore, be viewed not as a geometric curiosity, but as a deliberate design principle. It determines the total topological charge that must be accommodated while simultaneously shaping the energetic pathways through which that accommodation occurs. In soft matter, this combination is especially powerful because a single confinement geometry can support several distinct realizations—singular or nonsingular, twisted or untwisted, linked or unlinked—depending on material parameters and the characteristic length scale. Confinement thus does more than reveal topology: it organizes the landscape on which competing topological states are selected.

This perspective remains central as we transition to a different physical setting. In droplets, the confining surface encloses the ordered medium itself; for embedded colloidal inclusions, the distorted director field occupies the surrounding nematic bulk. The underlying logic, however, remains invariant. In both cases, the genus and boundary topology of the surface constrain the **integral topological charge**, while local curvature and anchoring determine how that charge is manifested within the actual texture. The distinction is therefore not conceptual, but geometric: the same design principle is transferred from the curved confinement of an ordered medium to curved and topologically nontrivial objects embedded within it. This means that the same genus can be used to realize different experimentally addressable states by changing anchoring, elastic anisotropy, size, or thermal history.

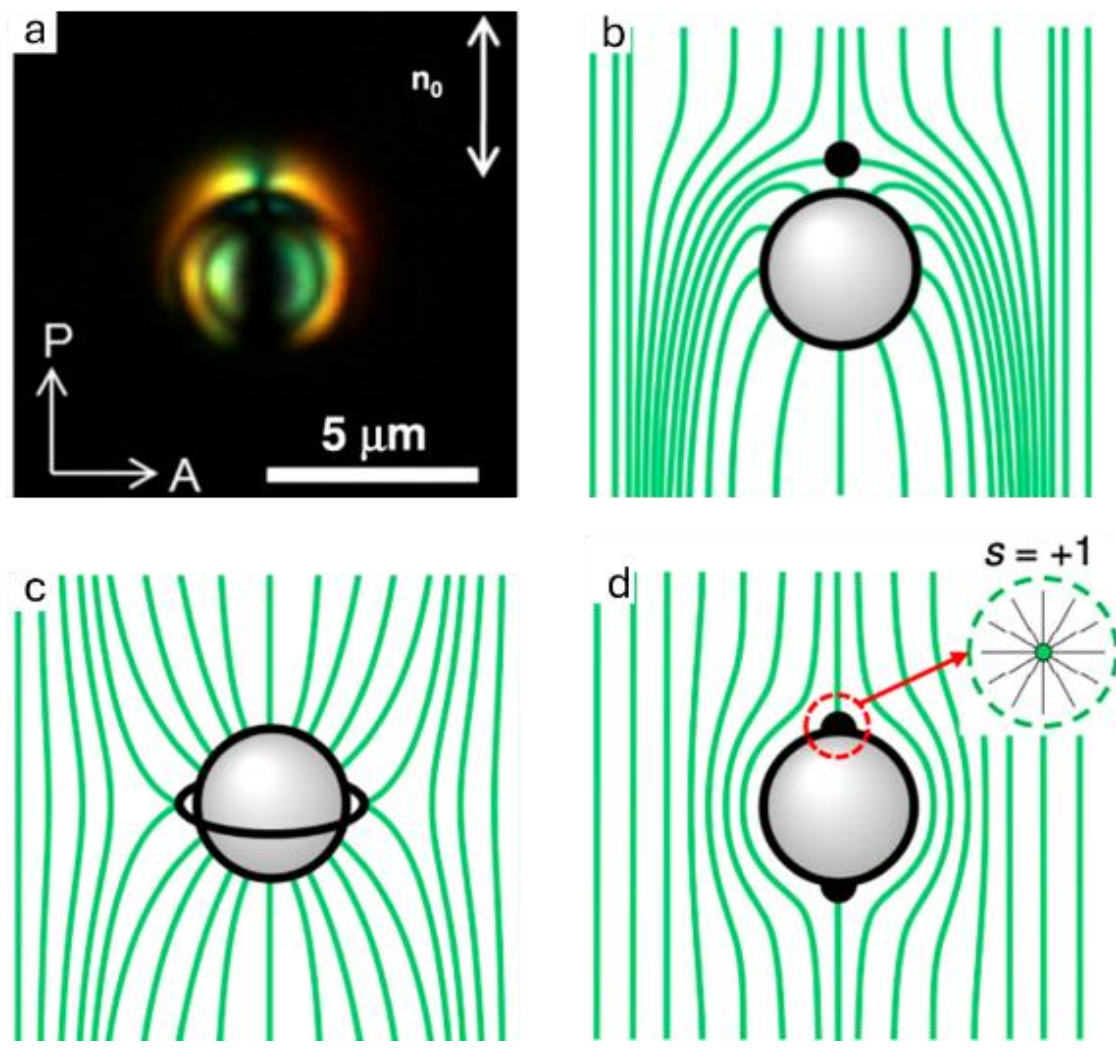

**Figure 7. Spherical colloidal particles with induced topological defects**. a, POM micrographs of an elastic dipole formed by spherical colloidal particles with homeotropic anchoring. Orientations of the polarizer and analyzer are labeled 'P' and 'A', respectively. b, Schematic illustration of the director field $\mathbf{n}(\mathbf{r})$ of dipolar symmetry at far-field, with a compensating hyperbolic point defect, black solid circles. c and d, Schematic representation of the $\mathbf{n}(\mathbf{r})$ featuring a quadrupolar symmetry at far-field and with homeotropic (c) and tangential (d) anchoring. The black ring in c represents Saturn ring half-integer disclination line. Boojums in d are marked as black-filled hemi-circles at the particle's poles along the far-field alignment. The inset in d shows the surface director field corresponding to each of the two boojums at the nematic-particle interface.

## 4. Colloidal inclusions with nontrivial topology

Colloidal inclusions with nontrivial topology extend the liquid-crystal colloid problem well beyond the familiar spherical case. Instead of asking how a particle with $\chi = 2$ perturbs an aligned nematic, one asks how the director field accommodates boundary conditions imposed by a surface of genus $g > 0$, by a surface with boundary, or, in more general theoretical settings, by a nonorientable surface. In such systems, the inclusion is more than a geometrical obstruction embedded in an ordered fluid. Its topology acts as a source of topological charge, defect structure, and elastic interaction. This is why topological colloids have become one of the clearest soft-matter realizations of the broader idea that geometry and topology can be used as design parameters for mesoscale organization [1,2,5].

The central issue is how the boundary conditions imposed by the colloidal surface are registered in the director field of the host nematic. For closed orientable inclusions, the global constraints are encoded in Eqs. (6)–(8): for tangential anchoring, the total winding number of the projected surface defects must add to the Euler characteristic $\chi$, whereas for homeotropic anchoring the effective hedgehog charge associated with the inclusion surface must be compensated by defects in the surrounding medium. These relations do not uniquely determine the resulting configuration, but they sharply restrict the admissible states. Once the topological requirement is fixed, local geometry, particle orientation, anchoring strength, and elastic relaxation determine how it is expressed in the actual texture [1,2,5,6].

It is instructive to read this section as a progression. One begins with the generalization from spheres to handlebodies, then moves to surfaces with boundary and sharp geometric features, and finally reaches the point where the topology of the inclusion becomes rich enough to lead naturally into knots and links. At each stage, the same question reappears in a slightly different form: how is the topology of the particle pre-determining the molecular alignment field and singular defects around the inclusion?

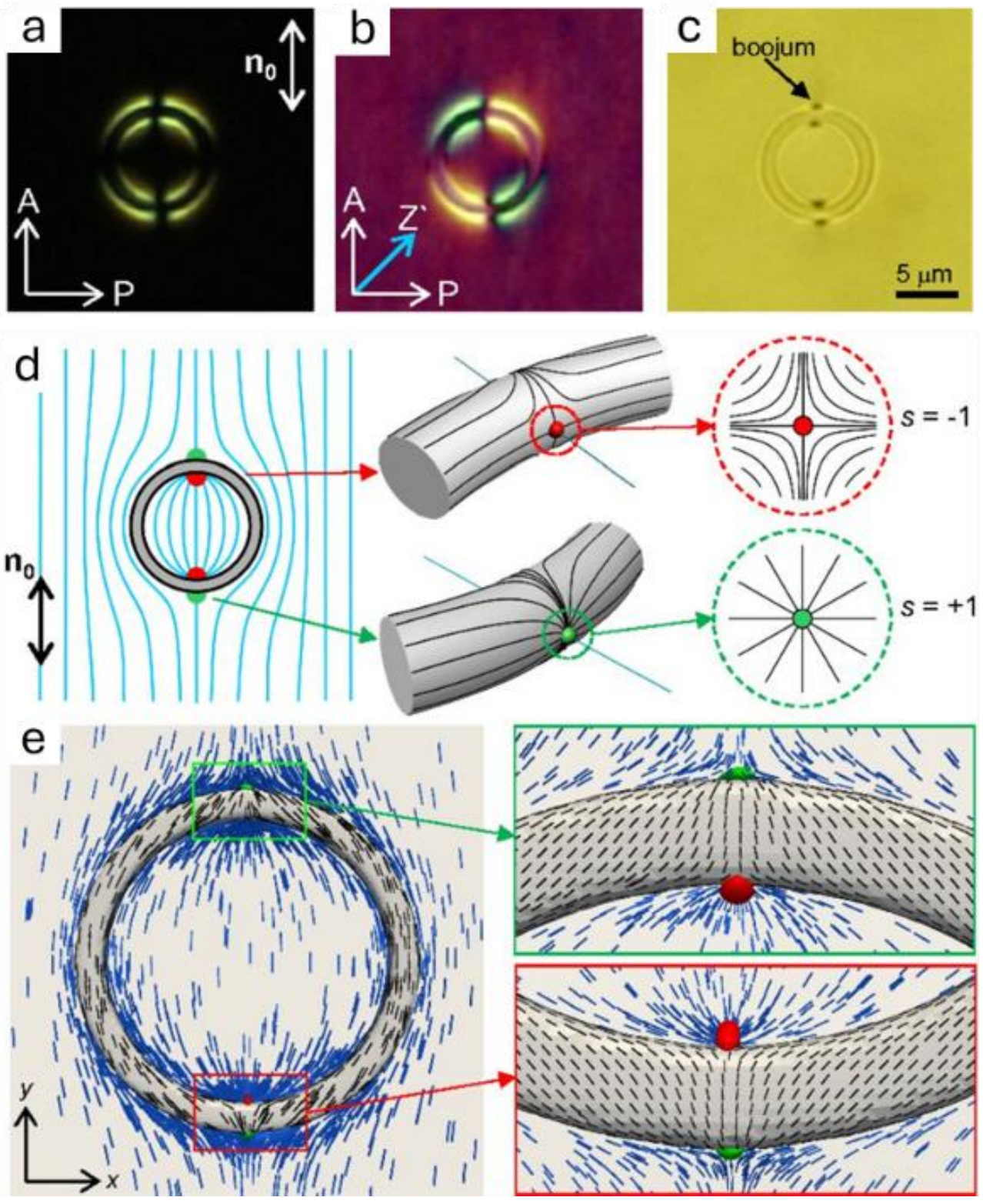

**Figure 8. $g = 1$ colloidal particles with tangential anchoring**. The colloidal torus is aligned along the far-field director. POM micrographs taken without (a) and with (b) a full-wavelength retardation plate with a slow axis $Z'$. c, Bright-field microscopy micrograph with 4 clearly visible boojums (dark spots). d, Schematic representation of director field around a torus aligned parallel to the far-field director $\mathbf{n}_0$. Insets focus on surface defects of the surface director field $\mathbf{n_s}(\mathbf{r})$. Green and red semispheres and circles represent, respectively, radial ($s = +1$) and hyperbolic ($s = -1$) surface disclinations in $\mathbf{n_s}(\mathbf{r})$. e, Numerically calculated $\mathbf{n}(\mathbf{r})$ in the bulk (blue rods) and $\mathbf{n_s}(\mathbf{r})$ on the surface (black rods) of a colloidal torus. Insets highlight the details of $\mathbf{n}(\mathbf{r})$ and $\mathbf{n_s}(\mathbf{r})$ near boojums and the isosurfaces of constant reduced scalar-order parameter are shown by green and red colours, respectively. Reproduced with permission from [6].

### 4.1. From spheres to handlebodies

The transition from ordinary colloids to topological colloids is most transparent when one moves from spheres to tori and higher-genus handlebodies. Spherical particles already illustrate the basic compensation principle. Under homeotropic anchoring, the radial director field on the surface behaves like a unit hedgehog and is compensated by a bulk point defect (figures 7a and b) or by a Saturn-ring disclination line (figure 7c). Under tangential anchoring, the projected surface director carries total winding number $+2$, most commonly realized by two boojums, figure 7d. Higher-genus inclusions preserve this general logic, but now the topology of the surface changes the net topological requirement from the outset [1,2,5,6].

For a handlebody of genus $g$, the Euler characteristic is $\chi = 2 - 2g$, so every added handle changes the net surface or bulk charge that must be accommodated. This is the central idea behind the topological-colloid framework [5]. In practice, particles in the forms of rings, multi-handle bodies, and related structures can be fabricated by photolithography, two-photon polymerization, and related methods, then surface-functionalized to impose homeotropic, tangential, or tilted anchoring [1,2,5]. Once dispersed in a nematic host, they provide a particularly clean setting in which to test how topological constraints derived from $\chi$ are projected onto real three-dimensional director fields.

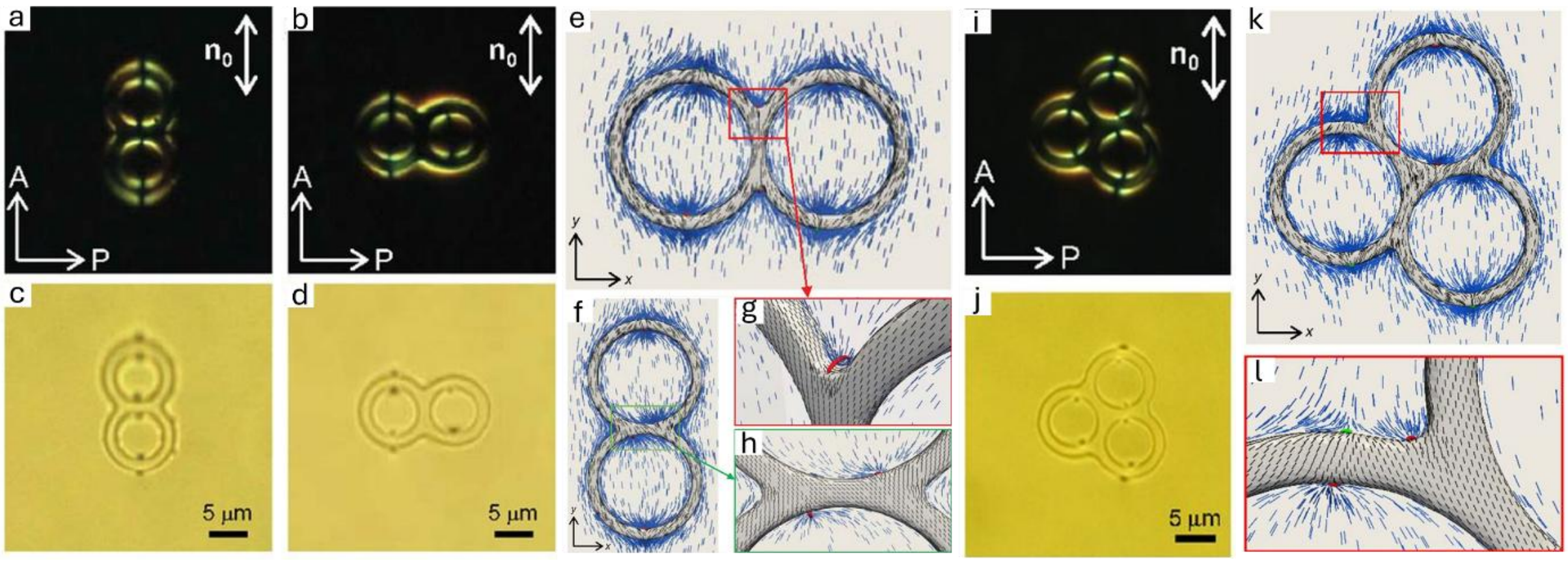

**Figure 9. $g = 2, 3$ colloidal particles with tangential anchoring**. POM (a,b) and bright-field (c,d) images of colloidal handlebodies having ring planes parallel to $\mathbf{n}_0$ and with the axis connecting the ring centres at different orientations with respect to $\mathbf{n}_0$. e and f, Numerically calculated LC bulk configurations (blue rods) and the surface director fields $\mathbf{n}_\mathbf{s}(\mathbf{r})$ (black rods) for the $g = 2$ particles for two different orientations relative to $\mathbf{n}_0$. g and h, detailed views of $\mathbf{n}(\mathbf{r})$ and $\mathbf{n}_\mathbf{s}(\mathbf{r})$ in the vicinity of the boojum regions marked in e and f, with the isosurfaces of constant scalar-order parameter shown in red. POM (i) and bright-field (j) images of a $g = 3$ particle; k, a numerically calculated LC bulk configuration (blue rods) and the surface director field $\mathbf{n}_\mathbf{s}(\mathbf{r})$ (black rods) for the colloidal handlebody. l, a detailed view of $\mathbf{n}(\mathbf{r})$ and $\mathbf{n}_\mathbf{s}(\mathbf{r})$ in the near-boojum region marked by the red rectangle in k, the green and red tubes correspond to isosurfaces of constant reduced scalar-order parameter. Reproduced with permission from [6].

What makes these systems instructive is that global topology constrains the total charge but leaves room for multiple textures. Even for a fixed genus, several stable or metastable configurations may arise, depending on particle orientation relative to $\mathbf{n}_0$, particle aspect ratio, anchoring strength, and the relative costs of splay, twist, and bend. Thus, a single toroidal particle can support more than one reproducible defect structure under otherwise similar conditions [1,5,6]. The topological class of the inclusion therefore constrains the solution space, while geometry and elasticity determine which specific structural realization actually occurs.

### 4.2. Tangential anchoring: boojums on colloidal handlebodies

The tangentially anchored case provides one of the most direct and vivid demonstrations of Eq. (6). Because the director projected onto the colloidal surface must remain tangent to that surface, the sum of winding numbers of the surface defects is constrained to equal $\chi$. For a torus, $\chi = 0$, so one may obtain either no boojums or self-compensating positive and negative boojums. For higher-genus handlebodies, the net winding number becomes increasingly negative, and the surface must accommodate a correspondingly larger total defect charge [1,6].

A central result in this area was the demonstration that handlebody-shaped particles with tangential anchoring indeed induce boojums whose total winding number follows the expected topological rule, while their detailed positions are controlled by free-energy minimization [6]. For a genus-one torus aligned with its ring plane parallel to $\mathbf{n}_0$, the most commonly observed configuration contains four boojums arranged in self-compensating pairs, even though the net surface charge remains zero (figure 8). For $g > 1$ handlebodies (figure 9), a larger number of boojums appears, with positions and multiplicities that may vary between stable and metastable states while the total winding remains fixed by topology. In addition to the topologically required negative defects, one often finds extra self-compensating boojum pairs, as well as split-core structures in which an integer-strength surface defect decomposes into two half-integer defects connected by a short handle-shaped bulk singular region. Although boojums of opposite sign may diffuse and annihilate over time as the director field relaxes between metastable and more stable configurations, the topological constraint on the total winding is always preserved. In practice, annihilation often stops before only the minimum topologically required number of boojums remains, because some of the surviving, but topology-redundant, defects still help lower the elastic free energy [6]. These observations show that boojums on topologically complex particles are not simply higher-genus analogs of the two boojums on a sphere; they belong to a considerably richer family of topology-compliant surface textures [1,6].

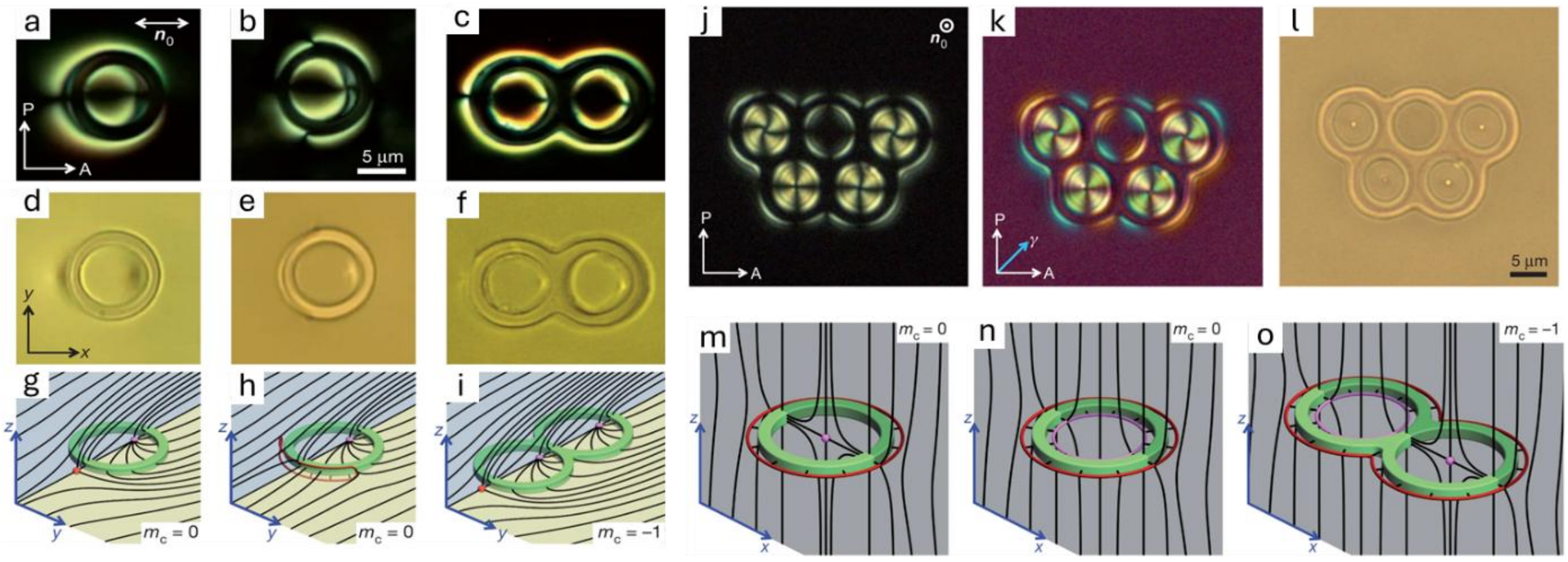

**Figure 10. Colloidal handlebodies with homeotropic anchoring**. a-i, Handlebody particles aligned with their ring planes parallel to the far-field director. POM (a-c) and bright-field (d-f) images of $g = 1,2$ particles; (g-i) show schematic director configurations corresponding to a-f. Magenta and red spheres in g-i represent hyperbolic point defects with the $m = +1$ and $m = -1$, while the closed red tube in h corresponds to a half-integer disclination ring with hedgehog charge $m = -1$. j-o show colloidal handlebodies oriented with their ring planes perpendicular to the far-field director. j–l, $g = 5$ handlebodies and corresponding $\mathbf{n}(\mathbf{r})$ visualized by POM without (j) and with (k) a retardation plate, and by bright-field microscopy (l). m-o, Schematic representations of typical $\mathbf{n}(\mathbf{r})$ induced by $g = 1,2$ handlebodies. Red and magenta lines represent outer and inner half-integer disclination line with the hedgehog charges $m = -1$ and $m = +1$, respectively; magenta spheres in the ring centres represent $m = +1$ hyperbolic hedgehog point defects. Reproduced with permission from [5].

An important lesson from these experiments is that the minimum topological requirement does not necessarily coincide with the energetic ground state. The surface may generate additional self-compensating boojum pairs if doing so lowers the cost of bulk elastic distortions. In this sense, tangentially anchored handlebodies provide a clear example of a broader principle: topology constrains what is allowed, but elasticity determines what is favorable. The resulting landscape of stable and metastable states can often be explored experimentally by local laser-induced melting and subsequent quenching back into the nematic phase, where different configurations are found to occur with certain probabilities [1,6].

### 4.3. Homeotropic anchoring: topological charge compensation in the bulk

For homeotropic inclusions, the emphasis shifts from surface boojums to the bulk defects required to compensate the topological hedgehog charge associated with the colloidal surface. Equation (7) implies that a closed orientable inclusion of genus $g$ carries effective surface charge $\pm(1 - g)$, and Eq. (8) requires the total charge of the induced bulk defects to be the opposite, assuring the overall topological neutrality. This leads to a hierarchy of defect structures that differs qualitatively from the tangential case [1,5] (figure 10).

A key experimental result was that homeotropic handlebody colloids can spontaneously align either with their ring planes parallel (figures 10a-i) or perpendicular (figures 10j-o) to the far-field director $\mathbf{n}_0$, and that each orientation selects a different compensating defect structure [5]. In one geometry, a handlebody may induce point defects inside and outside the handles (figures 10a, c, and i); in another, the same topological object may be accompanied by disclination loops in the holes of a handlebody or wrapping around the exterior (figure 10j). Disclination loops in the holes of a handlebody can be converted into point defects, and vice versa, by locally melting the liquid crystal with optical tweezers at laser powers above 100 mW and then quenching it back into the nematic phase [5]. This reversibility indicates that the corresponding director configurations have comparable free energies, albeit there is an energetic barrier preventing the transformation process to occur spontaneously, under the effects of thermal fluctuations. Colloidal particles with their ring planes parallel to $\mathbf{n}_0$ tend to induce point defects both inside the holes and near the particle exterior (figure 10g-i). These point defects can sometimes transform into disclination loops that follow the curved edge faces of the particles (figure 10h) while carrying the same effective topological charge as the point defects they replace. What remains invariant is not the specific morphology of the defects, but the total compensated charge. This is precisely the content of Eq. (8): it functions as a robust topological relation even when the actual defects are structurally diverse. Representative director configurations for handlebody colloids under homeotropic anchoring and at different orientations with respect to the far-field director $\mathbf{n}_0$ are summarised in figure 10.

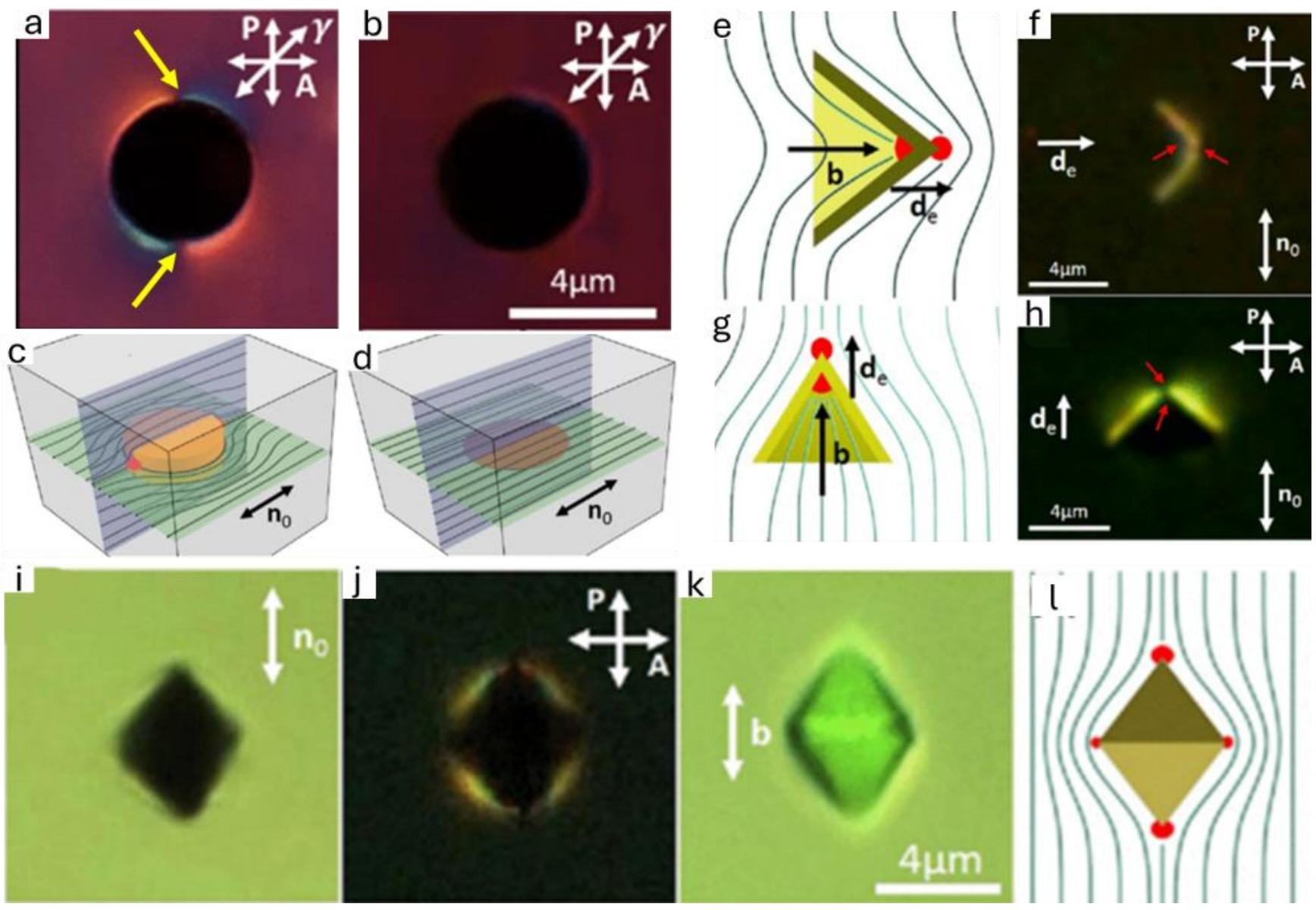

**Figure 11. Colloidal particles as orientable 2D manifolds with boundaries.** POM micrographs of colloidal plates made of a thick foil with the thickness $\approx 1$ $\mu$m (a) and a thin foil with the thickness $\approx 100$nm taken with a 530 nm phase retardation plate with a fast axis $\gamma$ inserted between the crossed polarizer ($P$) and analyzer ($A$); yellow arrows in a indicate boojums. c and d, Schematic illustrations of the director field around the thick (c) and thin (d) plates. The locations of the boojums on the surface of the thick plates are indicated by red hemispheres in c. The thin plate in d interacts with the $\mathbf{n}(\mathbf{r})$ as an orientable 2D manifold with a boundary, for which the Poincaré-Hopf theorem is not applicable, and as a result, thin plates generate no boojums. e, schematic of $\mathbf{n}(\mathbf{r})$ and defects generated by a colloidal pyramidal cone fabricated from a thin foil and having a boundary. The orientation of the particle is specified by $\mathbf{b}$ which in this case is perpendicular to the far-field director; the experimental realisation of this configuration is presented in f, using POM. Vector $\mathbf{d}_e$ represents schematically elastic dipole moment imparted to the director field by the colloidal particle. g and h, show the colloidal pyramidal cone, whose $\mathbf{b}$ is parallel to the far-field director. Complementary red fragments of spheres in e and g show fractional boojums; and the orientations of crossed polarizer ($P$) and analyzer ($A$) and $\mathbf{n}_0$ are shown using double arrows. In f and h, the thickness of the foils is $\approx 100$ $nm$ and the red arrows point to the locations of boojums. i, Bright-field, j, polarizing, and k, reflection optical micrographs showing a colloidal octahedron formed through the assembly of two pyramidal cones in a nematic LC, with a schematic $\mathbf{n}(\mathbf{r})$ and boojums (red spheres) depicted in l. Reproduced with permission from [7].

Figures 8-10 collectively, emphasize how increasing genus reshapes the number, arrangement, and character of particle-induced surface boojums and bulk compensating defects, and thus make visible the distinct ways in which the same topological requirement is realized under different anchoring conditions. Equally important is the observation that particles often induce more individual defects than the topological minimum would appear

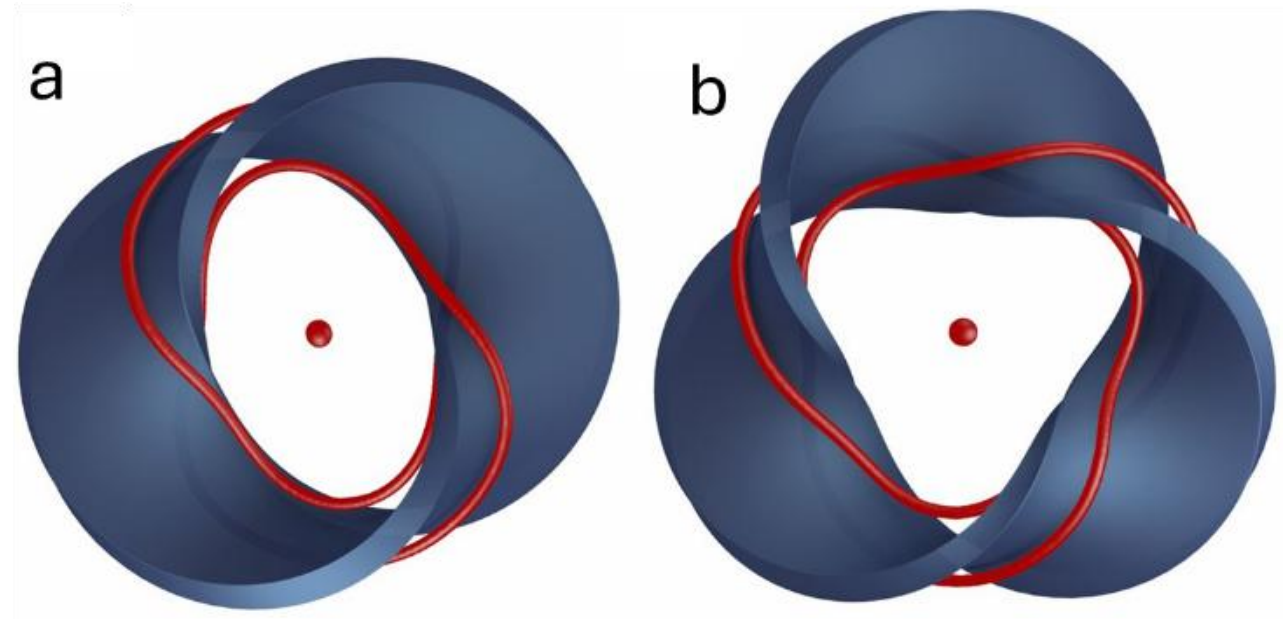

**Figure 12. Multiply twisted Möbius-strip-like inclusions in a chiral LC.** Linked (a) and knotted (b) disclination lines in a chiral LC generated by Möbius-strip-like colloidal particles with homeotropic anchoring on the wide flat surfaces and tangential on the narrow faces. a, particle with $p = 2$ which is accompanied by Hopf link (red tubes), and b shows p=3 particle with trefoil knot (red tube). The red small spheres in the centers show hedgehogs which appear in pairs above and below the strip. Reproduced with permission from [18].

to require. This happens because additional self-compensating defect pairs may reduce the total free energy by relaxing severe bend or splay distortions that would otherwise be needed to match the surface boundary conditions to the far field. Thus, for homeotropic handlebodies, the net charge is fixed, whereas the detailed texture remains to be determined. This distinction becomes relevant for self-assembly, because the elastic multipole character of the distortion field around the inclusion depends sensitively on the detailed defect arrangement, not only on the net total topological charge [2,5,17]. This point also makes the connection to materials design especially clear. A topologically nontrivial particle does not simply perturb the director locally; it produces a long-ranged, anisotropic elastic field whose symmetry reflects the defect structure governed by particle geometry and anchoring. As a result, topological colloids can behave as highly programmable elastic multipoles, with interaction channels that are inaccessible to ordinary spherical colloids in isotropic fluids [2,5,17]. This provides a direct design handle: by choosing genus, anchoring, and particle orientation, one can tune not only the defect content but also the symmetry of the long-ranged colloidal interaction.

### 4.4. Surfaces with boundary: thin foils, pyramidal cones, and apex boojums

Not all interesting colloidal objects are closed surfaces. A major conceptual advance was the recognition that sufficiently thin foil particles in a nematic can behave as **surfaces with boundary**, rather than as closed orientable particles obeying the usual Euler-characteristic constraints for closed surfaces [7]. This distinction is physically important because it introduces colloidal analogs of mathematical surfaces with boundary, whose coupling to the director field is not constrained in quite the same way as that of ordinary closed particles (figures 11a-d).

A particularly illuminating case is provided by thin foil discs with tangential anchoring (figures 11b and d). When their thickness is much smaller than the extrapolation length, they can effectively induce no defects at all, behaving as orientable surfaces with boundary rather than as conventional closed colloidal objects [7]. Yet when such physically thin surfaces are shaped into hollow pyramidal cones, their geometry generates localized defects at the apex points (figures 11e-h). These apex boojums have fractional hedgehog character and occur in self-compensating pairs associated with the inner and outer pyramidal surfaces. When two such pyramidal cones self-assemble into an octahedral object, the open surfaces with boundary are converted into a closed surface without boundary (figures 11i-l), and the resulting defect content becomes consistent with the usual topological rule for closed surfaces [1,7].

These systems are notably relevant to the themes discussed here because they show that geometry can be decisive even when topology is formally simple. The sharp apexes and edges of the foil particles provide geometric singularities that pin or nucleate defects, while the presence of a boundary component changes the topological class of the inclusion itself. In other words, the liquid-crystal response can be tuned not only by changing genus, but also by changing whether a surface is open or closed. This substantially enlarges the design space available for mesoscale self-assembly [7].

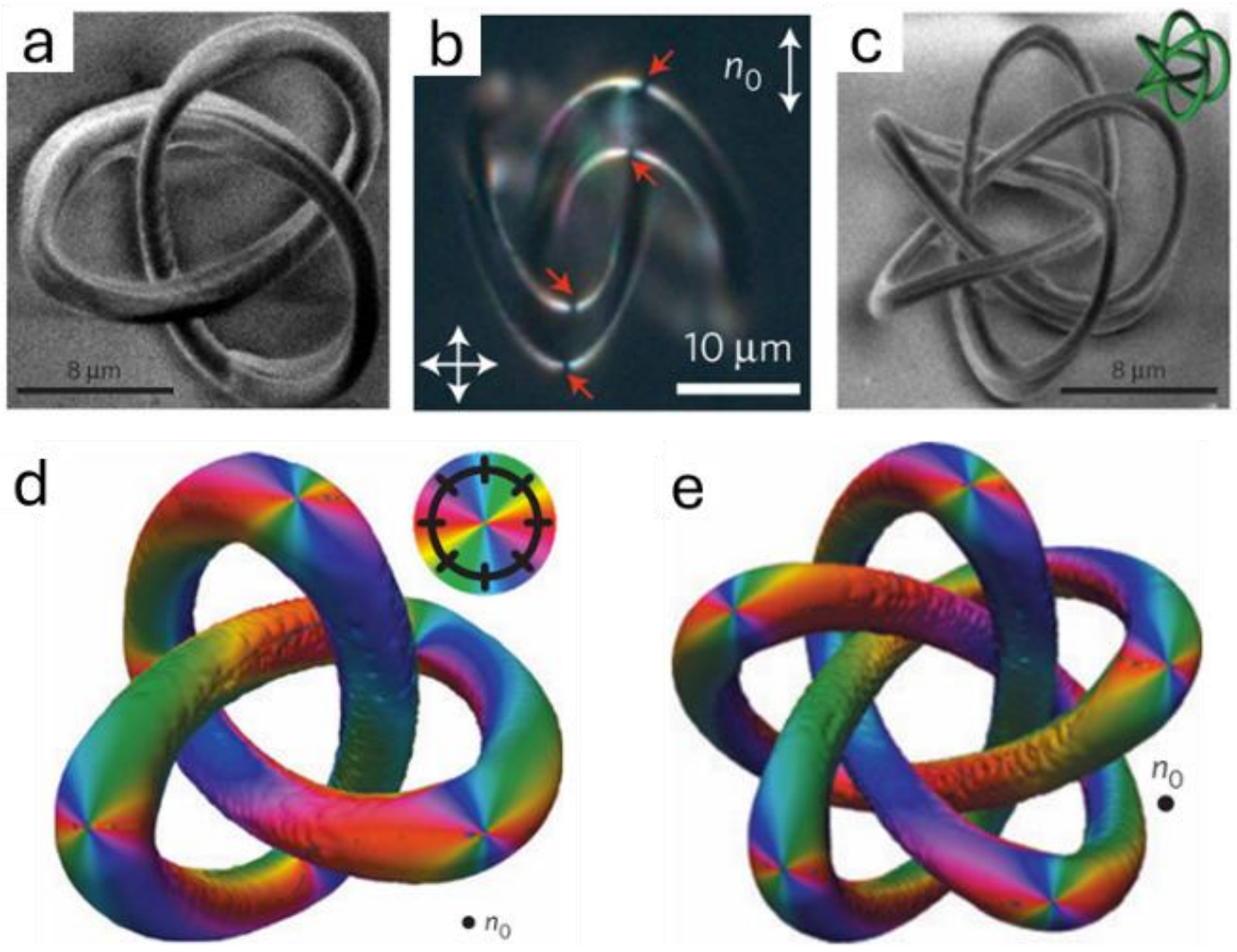

**Figure 13. Knotted colloidal particles with tangential anchoring.** a, Scanning electron micrograph of a T(3,2) knot shown as viewed along the torus axis. b, POM micrographs taken between crossed polarizers and with an additional 530 nm retardation plate with its slow axis aligned as shown by the blue double arrow; orientations of crossed polarizers are shown by white double arrows, and the red arrows points to the locations of boojums. c, Scanning electron micrographs of a T(5,3) knot with the corresponding 3D model shown in the top-right corner by green tube. Reproduced with permission from [20]. Three-dimensional reconstruction of the surface director field induced by a T(3,2) (d) and T(5,3) (e) knots. Colour patterns shows the azimuthal orientation of $\mathbf{n}(\mathbf{r})$ when projected onto a plane orthogonal to $\mathbf{n}_0$ according to the colour legend in the inset in d. The points where different colours meet correspond to the boojum. Reproduced with permission from Ref. [22].

A related line of work concerns faceted colloids with sharp edges. Such particles can host edge-bound disclinations of fractional winding and mediate transformations between bulk defect segments and surface-pinned singular lines. Once again, the lesson is that once one departs from the geometry of a smooth sphere, the topology and geometry of the inclusion become inseparable from the topology of the induced field [1].

### 4.5. Beyond orientable closed surfaces

The examples discussed thus far focus on orientable inclusions, either closed or with boundary. A natural extension is to consider nonorientable surfaces and, more generally, ribbons whose embedding in three-dimensional space introduces additional topological complexity. One illustrative family of structures is obtained by taking a rectangular strip, applying $p$ half-twists, and identifying the short edges to form a closed band [18]. For odd $p$, the resulting surface is nonorientable and possesses the topology of a Möbius strip; for even $p$, the band is orientable and topologically equivalent to an annulus. In both cases, the extrinsic topology of the embedding is non-trivial, as the boundary traces a torus knot or link depending on $p$.

Disclination lines following such boundaries can thus inherit prescribed knot or link types, including Hopf-link and trefoil-knot configurations (figure 12). Chirality plays an essential role in stabilizing these structures: in achiral nematics, these knotted defects generally relax into simpler, untangled disclination loops, whereas chiral nematics can support and sustain these more complex configurations [18]. Although this direction is still in its early experimental stages compared to studies of orientable handlebodies, it points to a significant extension of the topological-colloid paradigm. In this framework, both the intrinsic surface topology and the extrinsic embedding of the inclusion are used to prescribe the topology of the associated defect lines [1, 18].

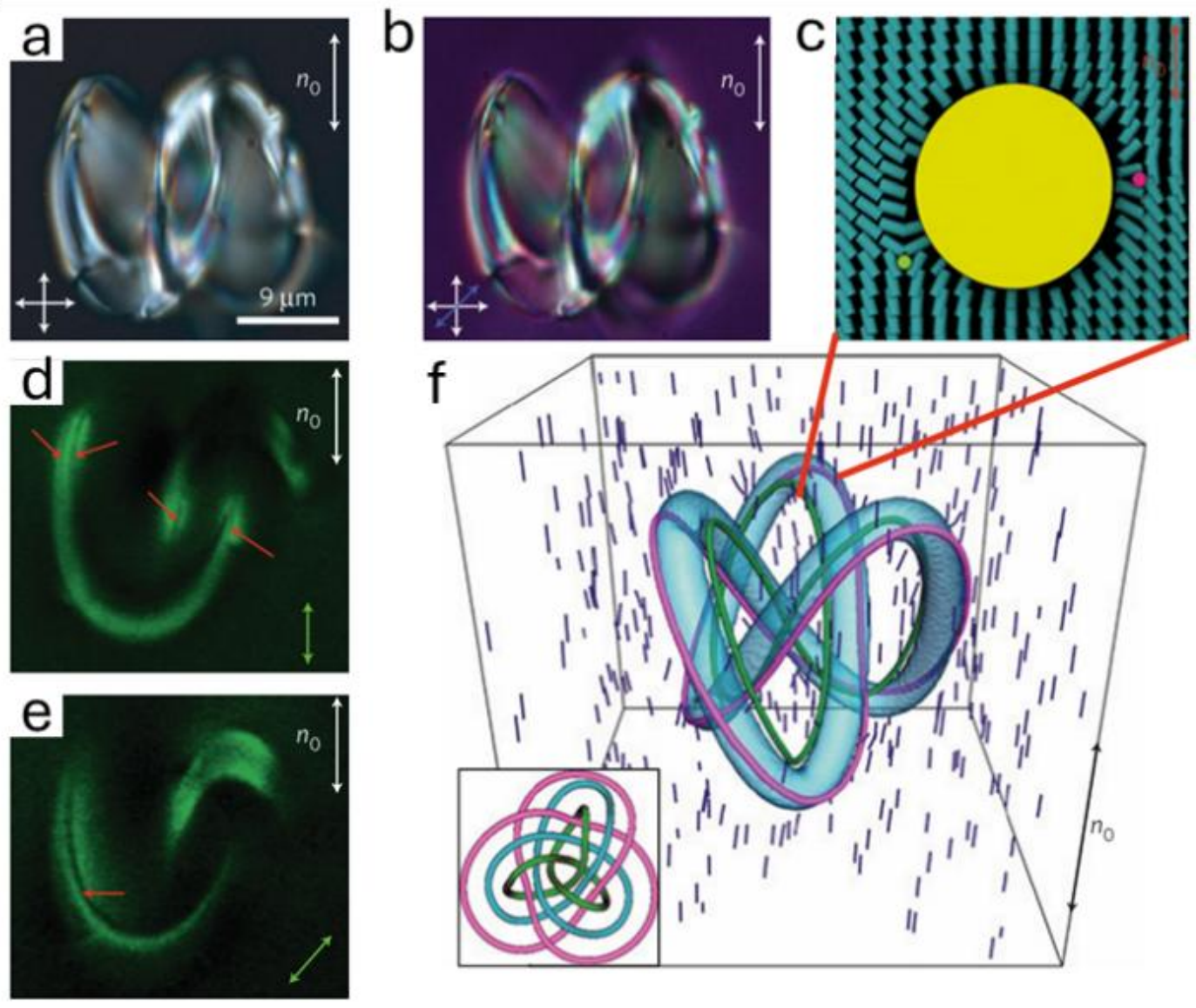

**Figure 14. Knotted colloidal particles with homeotropic surface anchoring.** LC configuration around a T(3,2) colloidal visualized by POM micrographs taken between crossed polarizers show by white double arrows in a, and with a 530 nm retardation plate shown by the blue double arrow in d. c, Numerically obtained $\mathbf{n}(\mathbf{r})$ calculated numerically and visualised by blue cylinders in a cross-section plane perpendicular to the knot tube, shown by transparent blue surface marked f. d and e, Show nonlinear polarized fluorescence images of $\mathbf{n}(\mathbf{r})$ corresponding to a and b, which are obtained at different orientations of the excitation-light polarizations (green double arrows) with respect to $\mathbf{n}_0$. Red arrows indicate the locations of the disclination lines. f, $\mathbf{n}(\mathbf{r})$ calculated numerically, the far-field director $\mathbf{n}_0$ is shown by black double arrow. Green and magenta tubes represent isosurfaces with a reduced scalar-order parameter, and represent knotted and mutually linked disclination lines. The bottom-left inset shows a topological schematic of the mutual linking between the particle knot (blue) and defect knots (green and magenta). Reproduced with permission from [22].

From the standpoint of this review, the primary conclusion is that “colloidal inclusions with nontrivial topology” constitutes a broad and fertile category. It encompasses a diverse spectrum of objects—handlebodies, open surfaces with boundaries, faceted shells, and nonorientable ribbons—whose common feature is the ability to encode specific geometric and topological information into the surrounding director field. Once this encoding is understood, these objects become precision tools for programming both local defect morphology and far-field elastic interactions.

This provides a natural transition to the following section. If the topology of a colloidal surface can organize the surrounding defect structure, one may then consider the case where the inclusion itself is fabricated as a physical knot or link. In such systems, the particle is no longer characterized solely by genus and boundary components, but also by knotting and linking invariants, leading to liquid-crystal textures of even greater complexity [1].

## 5. Knots and links in colloids and defects

The preceding sections showed how the topology of a colloidal surface or a confining boundary constrains the director field through genus and Euler characteristic. Knots and links introduce a further layer of topological structure. A knot-shaped particle is not distinguished from a sphere or torus merely by its Euler characteristic: it is also characterized by a knot type, for example the trefoil or, more generally, a torus knot $T(p,q)$, while multi-component particles require linking invariants as well. In nematic liquid crystals, these particle topologies can be expressed in the surrounding orientational field, so that the induced singular defects themselves become knotted or linked. This is one of the clearest examples of topology-guided design in soft matter, because the knot or link type carried by the colloid can be directly imprinted onto the induced defect structure [1].

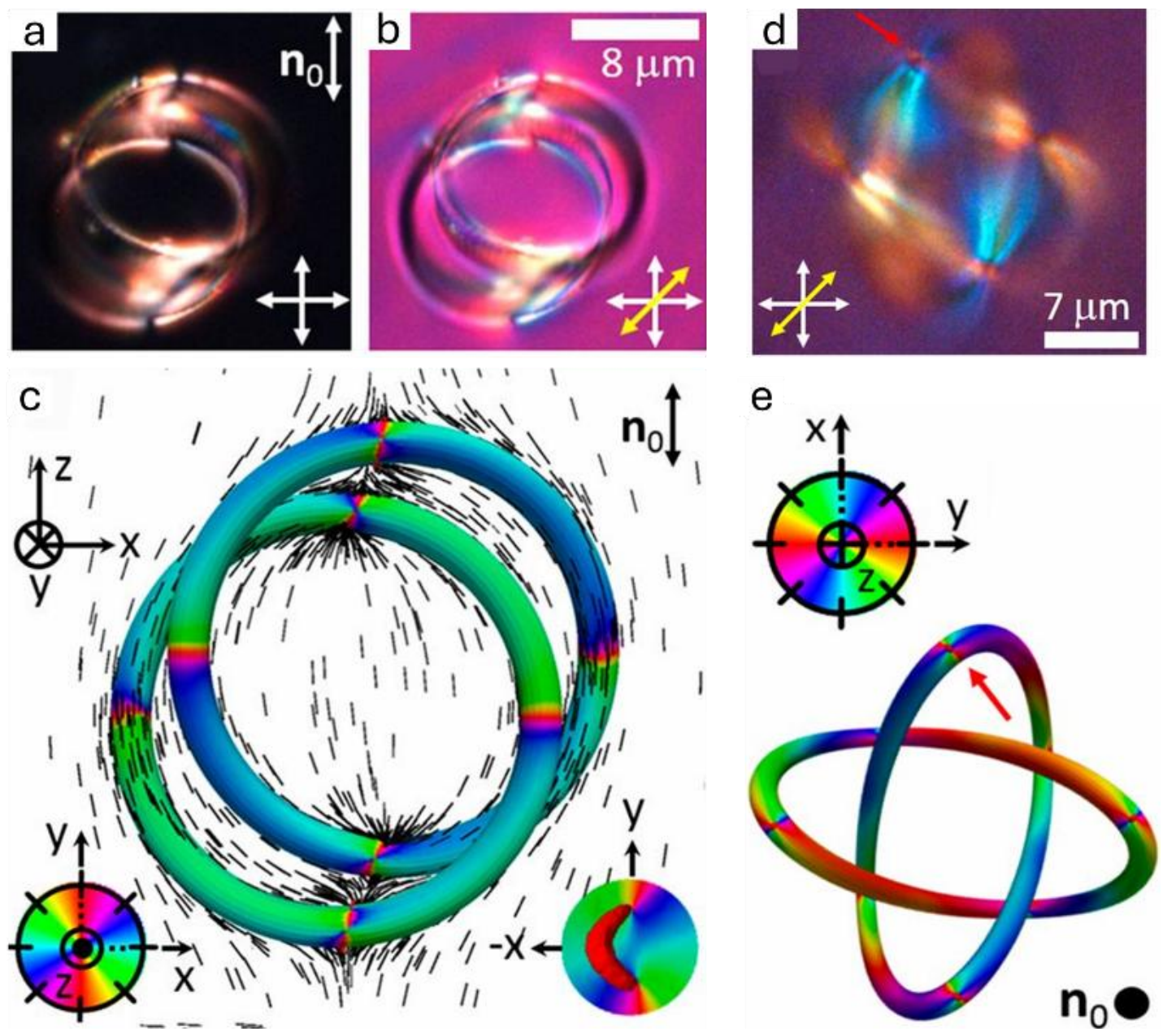

**Figure 15. Linked topological colloids with tangential anchoring.** POM micrographs of a colloidal Hopf link in a nematic LC obtained between crossed polarizers (white double arrows), a, and between crossed polarizers and an additional 530-nm wave plate (yellow double arrow) with a slow axis at 45° to the polarizer and analyser, b. c, Numerically calculated $\mathbf{n}(\mathbf{r})$ represented by black rods in the bulk and with colours on the surfaces of the linked tori, where the colour encodes the azimuthal orientation of $\mathbf{n}(\mathbf{r})$ with respect to $\mathbf{n}_0$ according to the colour legend in the lower-left inset. The lower-right inset, focuses on the details of the core structure of a boojum splitting into a semiloop of a half-integer defect line with the handle-shaped region of reduced order parameter shown in red. d, A Solomon link in a homeotropically aligned nematic cell as viewed between crossed polarizers aligned along the white arrows and an additional 530-nm wave plate with slow axis at 45° to them (yellow double arrow). e, A numerical result for the colour-coded surface director $\mathbf{n}_\mathbf{s}(\mathbf{r})$ where colours encode the azimuthal orientation of $\mathbf{n}(\mathbf{r})$ with respect to $\mathbf{n}_0$ according to the legend in the top-left inset. The Solomon link configuration corresponds to that shown in d. Reproduced with permission from [23].

The significance of this development is twofold. First, it extends the idea of topological colloids from particles classified only by genus to particles classified also by knotting and linking. Second, it shows that the topology of the singular structure is not limited to the geometry of the particle alone. Under appropriate boundary conditions and confinement, the surrounding nematic can sustain knotted and linked disclinations that may be interlinked with the colloid, linked only with one another, or, in some cases, appear even in the absence of a knotted inclusion [19-21]. Knotting and linking therefore amount to more than an expansion of the catalog of possible structures. They open a regime in which the topology of the field itself becomes an experimentally controllable degree of freedom.

### 5.1. Knot-shaped colloidal inclusions

A natural step beyond handlebodies is to fabricate colloidal particles in the form of mathematical knots (figures 13a-c), with a well defined circular cross-section of the tube being knotted while tracing knotted geometric constructs from the pure math's branch of knot theory. This became possible through two-photon polymerization, which allowed the realization of torus-knot particles such as the trefoil, as well as more complicated knots, with controlled surface anchoring in a nematic host [1,22]. For these inclusions, the Euler characteristic of the colloidal surface remains $\chi = 0$, just as for a torus, because the particle surface is topologically equivalent to a closed tube. What changes is the embedding of that tube in three-dimensional space, and it is this embedding that profoundly alters the possible director configurations.

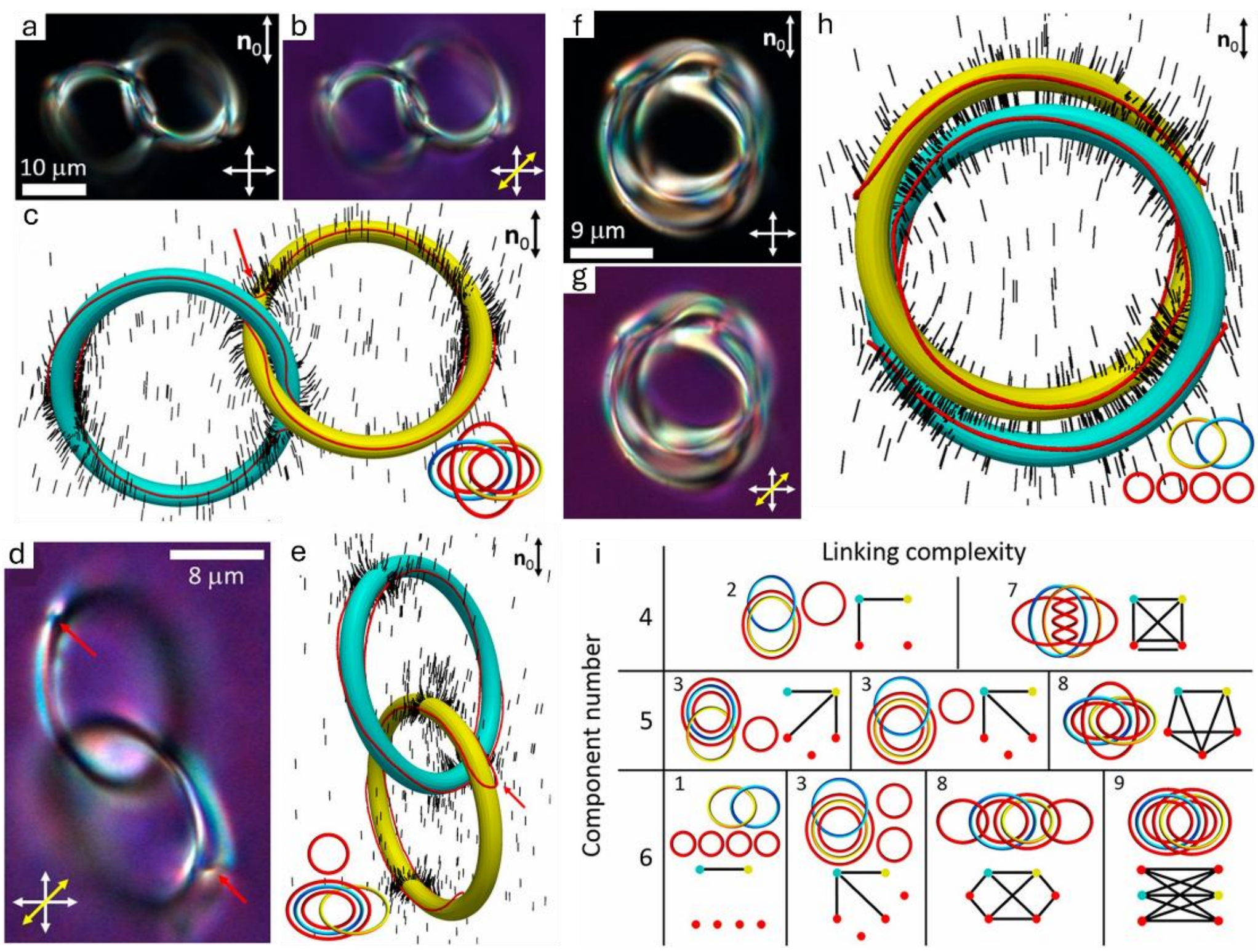

**Figure 16. Linked topological colloids with homeotropic anchoring.** POM micrographs of colloidal Hopf links with varying geometry in a nematic LC obtained between crossed polarizers (white double arrows) (a, f) and between crossed polarizers and an additional 530-nm wave plate (yellow double arrow) with a slow axis at 45° to the polarizer and analyser (b, d, g). Theoretical models of the experimental colloidal Hopf link configurations, with $\mathbf{n}(\mathbf{r})$ represented by black rods in the bulk and half-integer disclination lines represented by isosurfaces of reduced scalar order parameter (c, e, h); the insets depict the topology of linked colloidal components (blue and yellow) and disclination loops (red). i, Topology and graphical representations of interlinked Hopf link particles and accompanying closed defect loops. The linked colloidal tori are represented by blue and yellow rings, and the disclination line loops are shown by red rings. In the graphs, the individual links are indicated by black edges connecting the corresponding red-blue-yellow filled circles that represent colloidal or defect rings; the overall number of links is indicated next to the linking topology. All the presented cases were obtained based on the experimental data reported in [23]. Reproduced with permission from [23].

For tangential anchoring, knot-shaped particles generate boojums on their surfaces in self-compensating sets whose total winding number remains consistent with Eq. (6), (figures 13d and e). The key point is that, although the net surface charge still vanishes, the geometry of the knot determines where the boojums tend to appear and how the director winds along the colloidal tube. In the trefoil case, experiments and simulations showed that the stable state contains multiple boojums localized at geometrically selected regions of the knotted tube, with the overall pattern reflecting both the knot type of the particle and the far-field orientation $\mathbf{n}_0$ [1,22]. The particle is therefore not "topologically neutral" in any practical sense: even though the net charge of induced defects vanishes, meeting all known topological constraints related to the surface-field interaction at hand, the knot geometry imprints a highly structured field on the surrounding medium.

For homeotropic anchoring, the ensuing configurational consequences are even more striking. In that case, the knotted colloid induces singular defect lines that themselves become knotted. For a trefoil particle, as an example, the two main defect loops were found to be trefoils as well, mutually linked with one another and with the particle knot, so that the resulting object becomes a three-component defect-particle composite link [1,22] (figure 14). What this shows is that the colloidal knot does not simply carry topological information passively. It projects a corresponding topology onto the singular set of the surrounding field. Viewed this way, the nematic host acts as a medium that amplifies and externalizes the topology of the particle.

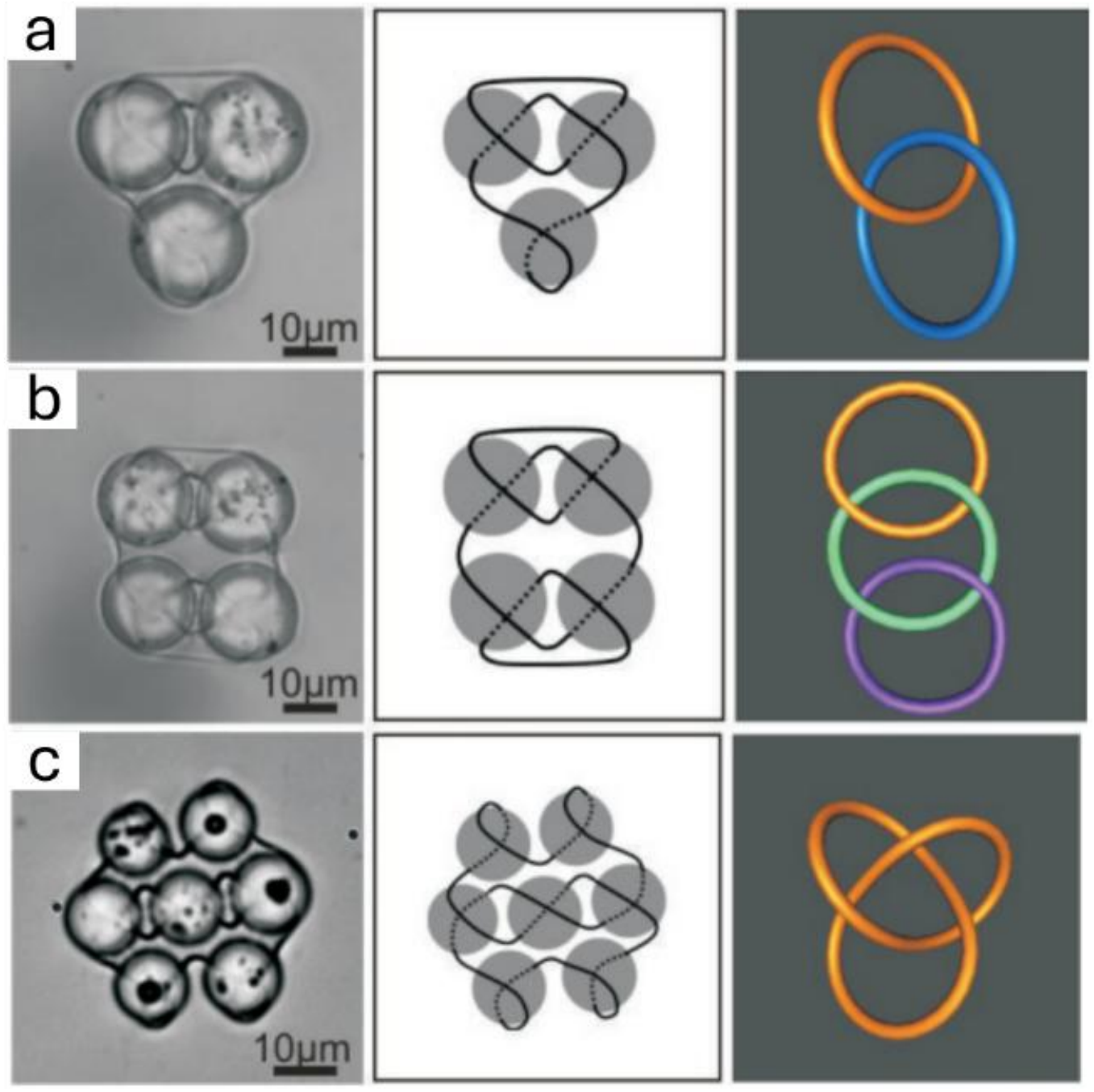

**Figure 17. Entangled colloidal clusters in a chiral nematic.** Clusters of spherical colloidal particles with homeotropic anchoring are entangled by disclination lines formed in right-handed π-twisted cholesteric LC cell. Left column shows POM images taken under crossed polarizers. The panels in the middle column represent schematics of the disclination loops, as deduced from images on the left. The right panels show the topologically minimized defect loop structure, after performing Reidemeister moves. Reproduced with permission from [20].

This is also the point at which topology begins to influence self-assembly through more than just elastic multipolarity alone. Knotted particles can interact not only through the anisotropic elastic distortions they induce, but also through the entanglement of defect loops that accompany them. This suggests routes to mesoscale architectures in which linking and knotting become effective interaction channels, supplementing more familiar defect- and orientational elasticity-mediated attraction or repulsion [1,22].

### 5.2. Linked composite colloids

A closely related development is the fabrication of colloidal particles whose topology is that of a multi-component link of closed loops, such as the Hopf link or the Solomon link, where again these colloidal mimics of mathematical constructs have finite cross-sections, usually circular with diameter in the submicrometer-to-micrometer range. In these objects the relevant invariants are no longer those of a single knotted loop alone, but also the mutual linking number of the components. Linked colloids have been fabricated and dispersed in a nematic host in such a way that their components remain mechanically connected while still undergoing Brownian motion relative to one another within the constraints of the link topology [23].

For linked particles with tangential anchoring, each ring component can induce boojums much as an isolated ring would, but the director field is now shaped by both the topology of the link and the elastic coupling between its components. Hopf-link-shaped colloids (figures 15a–c) most commonly exhibit configurations with eight surface boojums, four on each ring, while the ring planes are tilted with respect to the far-field director $\mathbf{n}_0$ [23]. These boojums occur in self-compensating pairs of opposite winding number, consistent with the zero Euler characteristic of each ring component, and may have split-core structures in the form of short half-integer defect-line semiloops. The elastic distortions around the linked rings weakly couple their positions and orientations, setting an equilibrium separation, an equilibrium crossing geometry, and a preferred angle relative to $\mathbf{n}_0$. Metastable configurations with different ring orientations and different numbers of boojums, including boojum-free states, can also occur [23]. In Solomon-link-shaped particles (figure 15d and e), the more complicated linking further enlarges the number of possible defect configurations and metastable states. Thus, once linking is introduced, the colloid acquires an internal topological degree of freedom that modifies the director field and the resulting colloidal interactions.

With homeotropic anchoring, linked colloids generate still more elaborate structures, in which the surrounding nematic hosts closed disclination loops. These defect lines may be associated with individual ring components or extend between them, effectively "jumping" from one component to another [23]. In addition to elastic distortions, the rings can become entangled by such line defect's closed loops, which act as elastic strings and couple their relative positions and orientations. As a result, a single Hopf-link particle can support multiple topologically distinct configurations of closed defect loops—linked with none, one, or both colloidal components, or with other loops—while remaining consistent with the overall topological constraints (figures 16a–h) [23]. The linked components remain elastically bound at preferred separations and orientations corresponding to local or global minima of the free energy. The result is a clear illustration of a recurring theme in this review: topology constrains the admissible class of configurations, while elasticity selects a particular realization within that class.

The diversity of homeotropic defect-loop states can be organized using simplified topological skeletons and graph representations, as shown in figure 16i. In Hopf-link colloids, two to four closed half-integer disclination loops may be linked with one ring, with both rings, with other defect loops, or remain unlinked. These mixed defect-colloidal links are topologically distinct and cannot be transformed continuously into one another unless defect loops cross, split, or merge, processes that involve substantial energetic barriers. The graph representation in figure 16i therefore provides a compact way to classify the resulting multicomponent links by the number of colloidal and defect-loop components and by the number of links between them. Even a single Hopf-link colloid can support long-lived metastable states with several defect loops and multiple mutual links, illustrating how the topology of the particles, the director field, and the singular defect network become intertwined [23]. Importantly, this summary of inter-linked multicomponent defect and colloidal ring structures is non-exhaustive but just illustrates a set of structures concurrently observed in both experimental observations and in computer simulations; hypothetically, a large variety of other configurations can emerge as well.

These linked composite colloids matter because they show that topological complexity can be distributed across several connected objects rather than concentrated in a single particle. The relevant "curvature effect" is no longer only the curvature of one individual surface, but the topology of an entire colloidal complex embedded in the nematic. That, in turn, suggests the possibility of using linked particles as mesoscale building blocks whose interaction rules are encoded not only by shape and anchoring, but also by linking number [23].

### 5.3. Knotted and linked defects without knotted particles

The examples above show how knotted or linked particles can template knotted or linked defect structures. A complementary route is possible when the particles themselves remain topologically simple, but the surrounding chiral nematic generates nontrivial defect topology. This occurs, for example, in chiral nematic colloids containing spherical particles or droplets under twisted boundary conditions, where defect loops formed around otherwise simple inclusions can organize into twisted loops, Hopf links, and trefoil knots [19,20]. As the number of inclusions increases, the defect topology can become progressively richer: colloidal trimers may be entangled by two closed defect loops forming a Hopf link (figure 17a), tetramers can support several mutually linked loops forming a short topological chain (figure 17b), and larger clusters can be threaded by a single knotted defect loop, such as a trefoil knot with three crossings (figure 17c) [20]. In these systems the particles themselves remain topologically simple, but the singular defect structure does not. The topology carried by the defects therefore need not be inherited directly from the colloids; it can instead arise from frustration in the chiral host medium, as well as stabilized by chirality or inclusions like multiple colloidal spheres.

This matters because it broadens the meaning of "knots and links in colloids and defects." The colloidal particle may be the primary source of the topology, as in knot-shaped inclusions, or it may merely act as a trigger or stabilizer for topologically nontrivial defects favoured by the chiral and confined nature of the host. From the standpoint of materials design, both routes are useful. The former employs particle topology as the primary programming tool; the latter uses confinement and chirality to write topological structures into the field even when the inclusions themselves are geometrically simple.

A related theoretical development showed that nonorientable surfaces in chiral nematics can enforce knotted defect structures [18]. This reinforces a broader principle already visible in the earlier sections: once the order parameter is coupled to a surface or confinement of nontrivial topology, the topology of the resulting singular set may be dictated as much by the global geometry of the boundary conditions as by the local structure of the particle itself. In other words, knotted and linked defects should not be viewed merely as special consequences of unusual complex shaped (like knot- or link-shaped) colloids; they are natural examples of closed-loop configurational outcomes of the coupled topology of field, surface, and confinement.

### 5.4. Why knots and links matter for self-assembly

The significance of knotting and linking extends far beyond mere classification. Knotted and linked defect loops possess finite line tension and interact elastically with other defects and inclusions, imposing constraints that are entirely absent in ordinary colloidal suspensions. In nematic soft matter, a pair of objects may be prevented from separating not only by elastic attraction, but also by the topological entanglement of the surrounding defect lines, which cannot be undone without energetically costly reconnection or additional defect generation and annihilation events. Knotting and linking therefore introduce a distinct topological interaction channel, fundamentally supplementing steric, electrostatic, magnetic, and depletion forces.

This becomes especially evident in linked colloids and chiral colloidal clusters. In linked particles, defect loops can act as elastic tethers that constrain the relative positions and orientations of the colloidal components. Conversely, in chiral nematic clusters assembled from topologically simple particles, the defect network itself can spontaneously form Hopf links, chains of linked loops, or trefoil knots. In both cases, the resulting assembly is governed not only by local elastic distortions, but by the global topology of the combined particle–defect architecture. Thus, knots and links are more than visual signatures of complexity; they are active components of the material's effective interaction landscape.

Liquid crystals are especially valuable here because knotted and linked structures can be created, imaged in three dimensions, and manipulated reversibly. Knot-shaped inclusions, linked colloids, and knotted defect loops therefore provide experimentally accessible realizations of a broader principle: field configurations in curved or frustrated media can carry higher-order topological invariants that go beyond local order and simple point-defect charge [1].

The discussion in this section also provides a natural transition to the next. Knotted and linked singular structures represent one route by which chirality and confinement generate nontrivial topology. Another route is through fully or largely nonsingular localized textures—such as skyrmions, torons, and hopfions—whose topology resides in the global organization of the director field rather than in singular lines. The passage from knots and links in disclinations to chiral localized textures is therefore not a change of topic, but a shift from singular to solitonic realizations of topological complexity [1,4]. In practical terms, knotting and linking therefore offer a route to assemblies whose stability is protected not only by energy minima, but also by topological constraints on how defects can reconnect or disentangle.

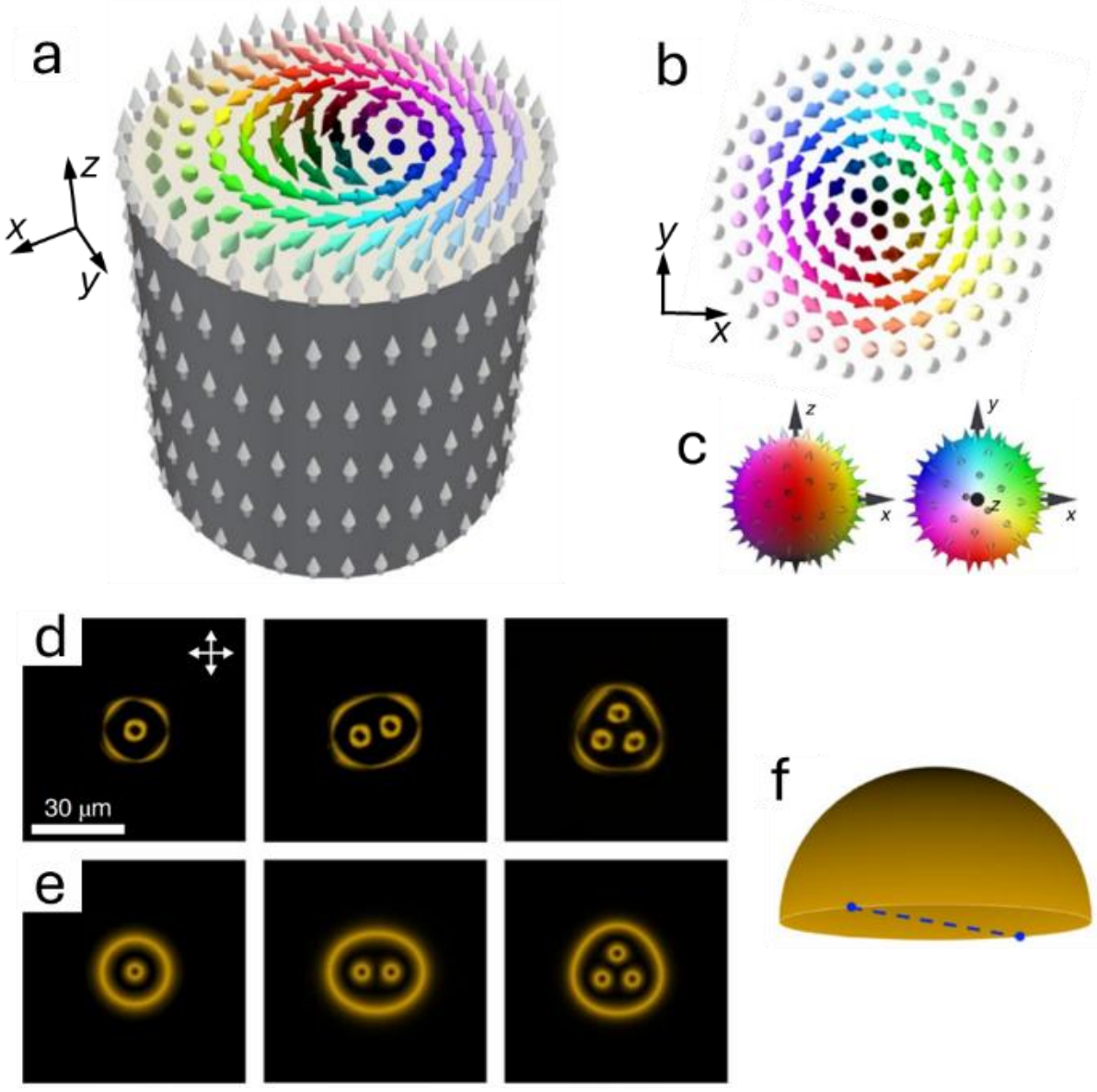

**Figure 18. Liquid crystal skyrmion and skyrmion bags.** A translationally invariant skyrmion tube (a) its projection onto the $(x, y)$ plane (b) visualized by arrows decorating $\mathbf{n}(\mathbf{r})$ smoothly and colour-coded according to the legends in c. POM micrographs of skyrmion bags ($N_{sk} = -1$) with one-to-three antiskyrmions ($N_{sk} = +1$) inside (d), and corresponding configurations calculated numerically. Colour corresponds to the $z$ – component of the director field, with the colour legend shown in f. Panel f schematically represents the director field manifold $RP^2$, where the blue dashed line connects the identical "configurations". Reproduced with permission from [36].

## 6. Chiral spatially localized textures: skyrmions, torons, and hopfions

Chirality adds a distinct layer to the interplay of curvature, topology, and confinement. In an achiral nematic, geometrically imposed frustration is typically accommodated through a combination of elastic distortions, singular defects, and escaped textures. In a chiral nematic medium, by contrast, the preferred local state already contains twist. Confinement can therefore do more than merely redistribute distortions: it can help select and stabilize spatially localized configurations that are topologically nontrivial even in the absence of embedded solid inclusions. At the same time, localized chiral solitons are not restricted to strongly confined geometries alone. In addition to thin chiral nematic films and cells, a variety of skyrmionic and related topological textures have been studied experimentally in thicker and more weakly confined systems [24,25], while theoretical work has explored broader stabilization mechanisms involving chirality, elasticity, saddle-splay, and external fields [26,27]. Confinement should therefore be viewed not as an absolute requirement, but as one particularly powerful mechanism for selecting, stabilizing, and controlling localized topological states. More generally, chirality turns curved or finite geometries into environments capable of supporting field-based excitations whose topology resides in the global organization of the director rather than solely in isolated defects [1,3,4].

From the perspective of this review, these states are important for two closely related reasons. First, they show that confinement and curvature do not act only by placing or reshaping singular defects at boundaries and inclusions; they can also help to stabilize bulk topologically nontrivial configurations that are partly or entirely nonsingular. Second, they provide a particularly vivid bridge from soft matter to the broader literature on skyrmions and hopfions in condensed matter and other physical systems. In liquid crystals, the corresponding order-parameter fields can often be reconstructed directly in real space, making them experimentally accessible realizations of structures that are otherwise difficult to probe in three dimensions [1]. The larger lesson is that defects, knots, and solitons are best viewed here not as disconnected categories, but as different ways in which geometry, topology, and chirality organize the same orientational field.

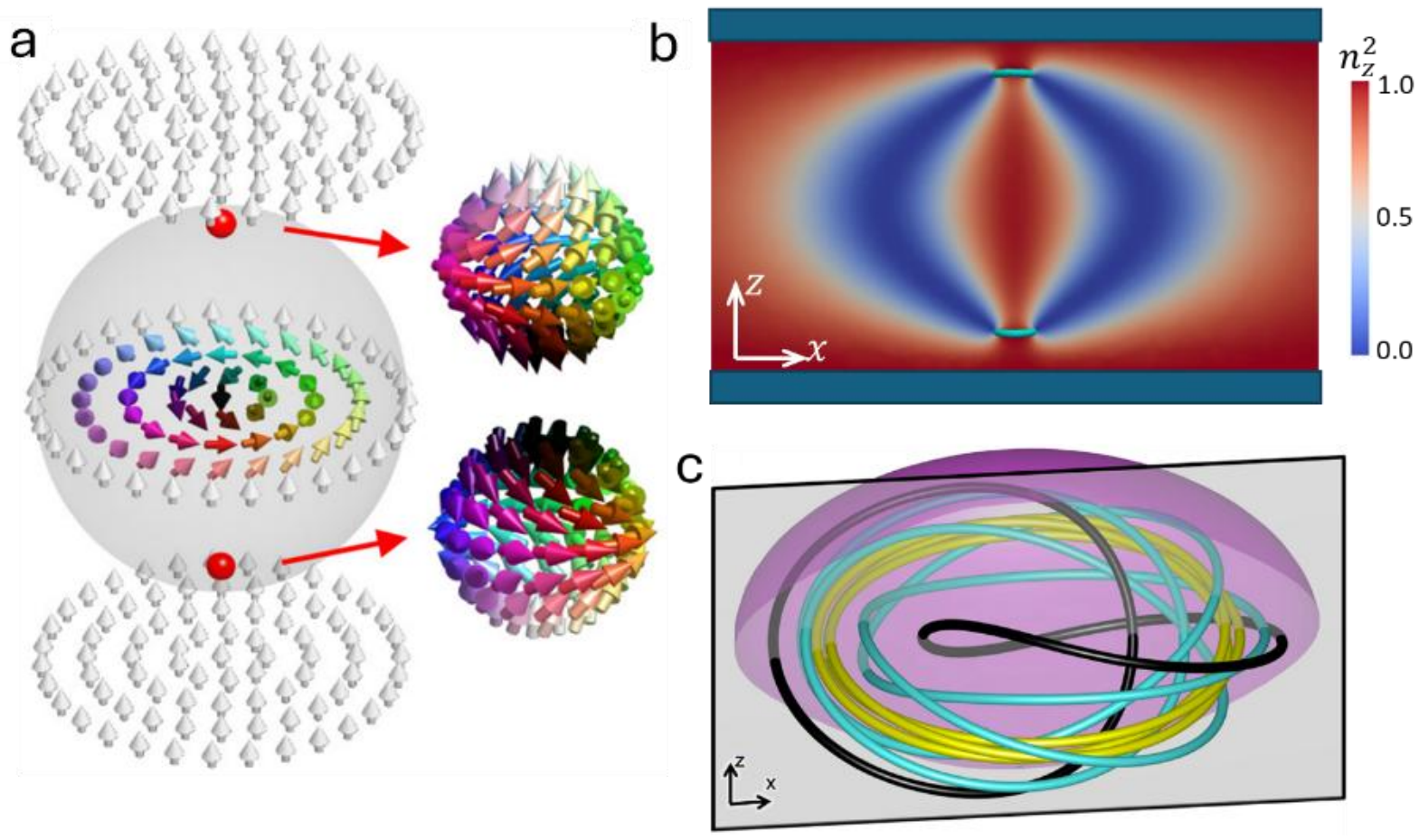

**Figure 19. Liquid crystal torons.** a, Schematic representation of the structure of a toron formed when a skyrmion tube (figure 18a) is embedded into a uniform director field formed, for example, in a homeotropic nematic cell with the thickness comparable with the cholesteric pitch. Confinement enforces termination of the skyrmion tube at point defects (small red spheres) – often called "Bloch points" in the terminology of magnetic systems; the detailed director configurations around the point defects are shown by coloured arrows. Colours encode points on a two-dimensional sphere as in figure 18c. Reproduced with permission from [26]. b, Cross-sectional view of the toron configuration as obtained by minimizing of the Landau-de Gennes free energy. Colour represents square of the $z$ −component of the director field, and blue tubes show half-integer disclination loops – open-core hyperbolic defects. c, Streamlines tangent to within the twisted region of a toron (outer magenta surface) that form a Hopf link (black tubes), pentafoil knot (cyan tube), and quatrefoil knot (yellow tube). Reproduced with permission from [37].

More broadly, the study of chiral topological solitons increasingly transcends the traditional boundaries between soft condensed matter, magnetism, and photonics. Torons, for example, were first realized in nonpolar chiral nematic liquid crystals, where the name itself was introduced, but related localized chiral textures were later identified in magnetic systems [28,29], in ferromagnetic chiral liquid-crystal colloids [4], and even in structured optical fields [30]. Likewise, hopfion-like structures in a helical background were first experimentally demonstrated in chiral nematic liquid crystals and later termed heliknotons [31,32]. Closely related three-dimensional solitons were subsequently predicted theoretically in chiral magnets [33] and have now been experimentally observed in magnetic solids such as FeGe [34]. These developments highlight an important broader point: chiral liquid crystals and related soft-matter systems provide experimentally accessible model environments for exploring the diversity of topological field configurations anticipated in magnetic materials and other ordered media. One especially promising future direction is the study of such solitons under intrinsically curved confinement, where chirality, topology, and geometric frustration can all become simultaneously active design parameters.

### 6.1. Two-dimensional skyrmions in confinement-frustrated chiral nematics

The simplest topological textures relevant to the present discussion are two-dimensional skyrmions. They arise when a chiral nematic is confined between substrates that favor nearly uniform alignment, while the bulk simultaneously prefers twist. Under suitable conditions, this competition is resolved not by uniformly twisting the entire sample, but by concentrating twist into localized double-twist regions embedded in an otherwise unwound background. In a vectorized description, such states are characterized by the skyrmion number

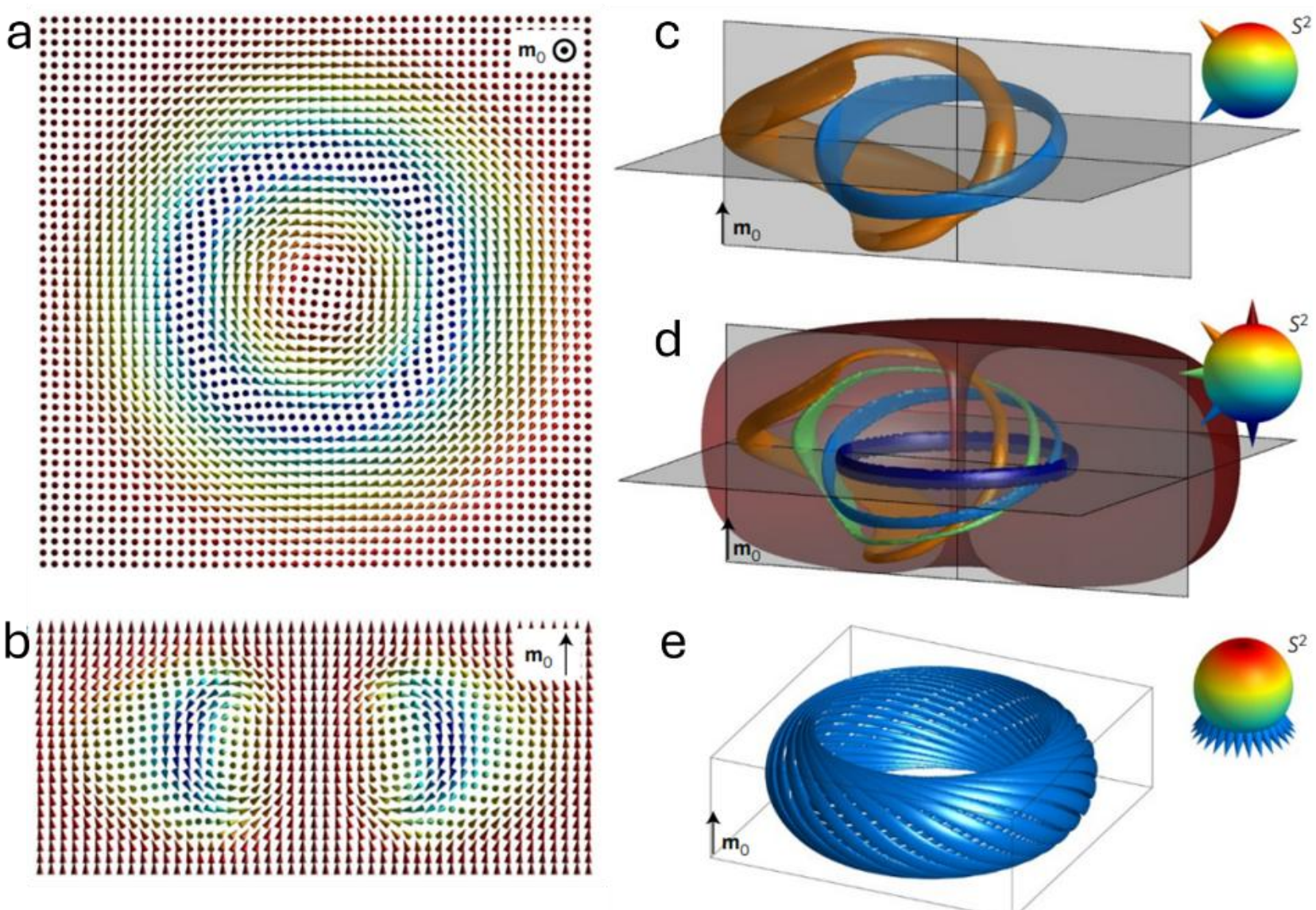

**Figure 20. $Q = 1$ hopfions in a chiral ferromagnetic colloidal fluid**. Numerically simulated magnetization profile $\mathbf{m}(\mathbf{r})$ of a chiral ferromagnetic colloidal fluid featuring $Q = 1$ hopfion configuration. a,b Cross-sectional views of the hopfion in the plane perpendicular/parallel to the far-field magnetization $\mathbf{m}_0$ (a/b). $\mathbf{m}(\mathbf{r})$ is shown using cones coloured according to the corresponding points on $S^2$ in the inset of c. Linking of preimages of the two (c) and five (d) representative points on $S^2$, including preimages of the south and north poles (the latter corresponds to $\mathbf{m}_0$ and only its half is shown for clarity). e, Torus-forming preimages of points on $S^2$ for constant polar but varying azimuthal orientations. Reproduced with permission from [4].

$$N_{\text{sk}} = \frac{1}{4\pi} \int \mathbf{m} \cdot \left( \partial_x \mathbf{m} \times \partial_y \mathbf{m} \right) dx\, dy, \tag{12}$$

where $\mathbf{m}$ denotes the suitably vectorized local orientation field in a plane normal to the skyrmion tube. For an elementary skyrmion, $N_{\text{sk}} = \pm 1$, whereas more elaborate composite configurations can carry larger integer values of this topological invariant.

What matters physically is that these states are not imposed by an inclusion and are not required by a topological charge placed at the boundary. They arise from the competition between two incompatible tendencies: confinement favors a nearly uniform director, whereas chirality favors twist. Full liquid-crystal skyrmions correspond to localized regions in which the director twists by $\pi$ from the center to the periphery, so that the cross-section covers the full order-parameter sphere after suitable vectorization (figures 18a-c). When embedded in a three-dimensional sample, such two-dimensional skyrmions form translationally invariant skyrmion tubes that are topologically distinct from the uniformly aligned background and cannot be removed without destroying the continuity of the order or introducing singular defects [35]. In this respect, skyrmions are among the clearest examples in soft matter of confinement and chirality stabilizing nontrivial topology. The same design principle was later extended to skyrmion bags, in which multiple antiskyrmions are enclosed within an outer skyrmion (figures 18d–f). These composite nonsingular textures provide access to skyrmionic states of higher topological degree, including degrees of either sign, depending on the number and nesting of the constituent skyrmions and antiskyrmions [36].

Although skyrmions are not “knots” in the ordinary geometric sense, they share with knotted structures the defining feature of topological robustness: they cannot be eliminated by continuous distortion of the field without loss of continuity or the appearance of singular states. They are thus most naturally viewed as the simplest chiral textures stabilized by confinement-induced frustration. In liquid crystals, their role goes beyond topological classification alone. They can assemble into arrays, interact elastically, and respond collectively to external fields, which makes them experimentally accessible mesoscale elements for organized structure formation [1,35,36].

### 6.2. Torons: finite three-dimensional twist in confined chiral media

A particularly important spatially localized state in chiral nematics is the toron [3]. A toron may be viewed as a finite-length skyrmion-like tube embedded in a homeotropic cell, with the twisted core terminated near the confining substrates by two point defects of opposite charge (figure 19a) [26]. In the vectorized description, the midplane cross-section maps onto $S^2$ once, as in an elementary skyrmion, while the two terminal point defects act as self-compensating elementary hedgehogs that connect this skyrmionic core to the uniform boundary-imposed background. Under certain conditions, the cores of these point defects can open into small half-integer disclination loops (figure 19b) [3]. Thus, although the cell midplane resembles a two-dimensional skyrmion, the full texture is intrinsically three-dimensional: it combines localized double twist, singular end caps, and an exterior region that remains topologically trivial [1,3].

Torons are especially significant because they provide one of the clearest examples of how confinement can transform geometric frustration into a stable topological state. They often appear when the intrinsic pitch of the chiral medium is incompatible with the nearly uniform alignment imposed by the confining surfaces, typically when the cell thickness is comparable to the cholesteric pitch. Under such conditions, the system lowers its free energy not by distributing twist uniformly throughout the sample, but by localizing it into mesoscale regions that satisfy the boundary conditions. The toron is therefore not simply an exotic defect configuration. It is a clear example of how finite confinement can create a stable topological state out of the balance between chirality, elasticity, and surface anchoring [1,3].

An important aspect of the toron is that it occupies an intermediate position between singular and nonsingular topology. On the one hand, the structure contains point defects near the boundaries and so is not fully nonsingular in the same sense as a hopfion. On the other hand, its internal structure is much richer than that of a skyrmion tube simply capped by two defects. Director streamlines inside a toron can form torus knots and links, including Hopf links, trefoils, and higher torus knots (figure 19c), depending on their position within the texture [1,37]. This occurs because the twist rate and local distortion pattern vary across the toron volume as the system embeds a three-dimensional twisted region into an otherwise uniform far field. Torons therefore connect the discussion of chiral solitons to the earlier discussion of knots and links: even when the singular set is relatively simple, the internal geometry of the orientational field may already carry a far more elaborate topological organization.

Torons are also important because they are experimentally robust and dynamically versatile. They can be generated optically, erased, translated, arranged into regular arrays, and driven collectively by electric fields. In other words, they behave not only as localized topological states, but also as particle-like mesoscale objects. This dual role—part field configuration, part assembly unit—is one reason torons form such a natural connection between the language of topology and the language of self-organization [3,38].

### 6.3. Hopfions: fully three-dimensional knot solitons

Torons therefore provide a natural bridge from skyrmion tubes to genuinely three-dimensional solitons. They already contain a finite twisted volume and can host knotted director streamlines, but their topology is still tied to singular termination points. Hopfions take the next step: the singular end caps are removed, and the topology is carried by the continuous three-dimensional field itself. This shift also connects liquid crystals to chiral magnetic systems. In confined noncentrosymmetric magnetic nanostructures, where the order parameter is a true vector magnetization rather than a nonpolar nematic director, Dzyaloshinskii–Moriya interactions, geometry, and perpendicular magnetic anisotropy at interfaces can stabilize localized hopfions with linked preimages and knotted emergent-field lines [29]. Surface anchoring in chiral nematics and interfacial anisotropy in magnets thus play closely analogous roles: both impose boundary conditions that help stabilize localized twisted regions against relaxation into the uniform background [29].

In a vector field, such as the magnetization field $\mathbf{m}(\mathbf{r})$ of a chiral ferromagnetic colloidal fluid, compactified physical space is mapped to the target sphere $S^2$. The corresponding topological class is labeled by the Hopf index, $Q \in \pi_3(S^2) = \mathbb{Z}$. Geometrically, $Q$ is most naturally understood through preimages: the set of spatial points at which the order parameter takes a given value on $S^2$. For a hopfion, the preimages of distinct points form closed loops in physical space (figures 20c and d), and the linking number of any two such loops equals the Hopf invariant. This interpretation is especially useful because it turns the abstract classification $S^3 \to S^2$ into a directly visualizable three-dimensional structure [1,4].

The experimental realization of hopfions in chiral magnetic colloidal fluids was therefore a major development [4]. In these systems, magnetic nanoplates dispersed in a chiral nematic host generate a fluid ferromagnetic medium with a genuine vector order parameter $\mathbf{m}(\mathbf{r})$. This makes the $S^2$ target space explicit and allows the linked preimages of different magnetization directions to be reconstructed experimentally. In this vector case, the topology is carried by the full nonsingular magnetization field rather than by isolated point or line defects. The resulting soliton occupies a finite three-dimensional volume, is embedded in a uniform far-field background, and is characterized by the linking of closed-loop preimages [1,4].

The liquid-crystal order parameter is the nonpolar director, $\mathbf{n}(\mathbf{r}) \equiv -\mathbf{n}(\mathbf{r})$, with order-parameter space $S^2/\mathbb{Z}_2 \simeq \mathbb{RP}^2$, rather than $S^2$. Nevertheless, suitable vectorization of $\mathbf{n}(\mathbf{r})$ allows one to assign a Hopf index and to analyze the topology through linked preimages. In experiment, such vectorization may correspond to doping the nonpolar chiral nematic with magnetically monodomain nanoplates that, under suitable experimental sample preparation, allow for decorating the director field with magnetization field [4].

The Hopf index invariant is encoded in the global linking structure of the order parameter field itself. This makes hopfions one of the clearest soft-matter realizations of curvature- and confinement-assisted topological solitons [1,4,39].

### 6.4. A unified view: from skyrmions to knot solitons

Although skyrmions, torons, and hopfions are often introduced separately, it is more useful here to view them as members of a broader family of confined chiral topological states. They differ in dimensionality, in whether singular defects are present, and in the order-parameter space used to classify them. Full skyrmions appear as localized two-dimensional textures or skyrmion tubes; torons are finite three-dimensional skyrmion-like textures terminated by point defects or small disclination loops; hopfions are fully three-dimensional nonsingular solitons characterized by linked preimages.

This unified viewpoint becomes sharper when torons and hopfions are interpreted through the geometry of preimages and director streamlines. In torons, the singular end defects complicate a direct Hopf classification, but the internal director structure can already contain closed streamlines forming torus knots and links. In hopfions, the corresponding topology is expressed more directly: preimages of order-parameter values form closed loops whose linking defines the Hopf invariant. Furthermore, we note that stable 3D heliknoton-type and hopfion-type Hopf solitons can be realized without confinement in thick samples and switched among different heliknoton-hopfion states by electric pulses, and can also coexist with geometrically different types of torons [25].

Seen in this way, the chiral states discussed here are not a digression from the earlier sections on droplets and inclusions. They are the bulk-field counterparts of those systems. In Sections 3–5, topology entered through the geometry of a confining surface or colloidal object. Here it enters through the geometry of the field itself, stabilized by confinement and chirality. In both settings, the same design logic is operative: geometry, topology, anchoring, and chirality constrain the admissible configurations, while elasticity selects the textures that can persist and, in many cases, behave as interacting mesoscale entities.

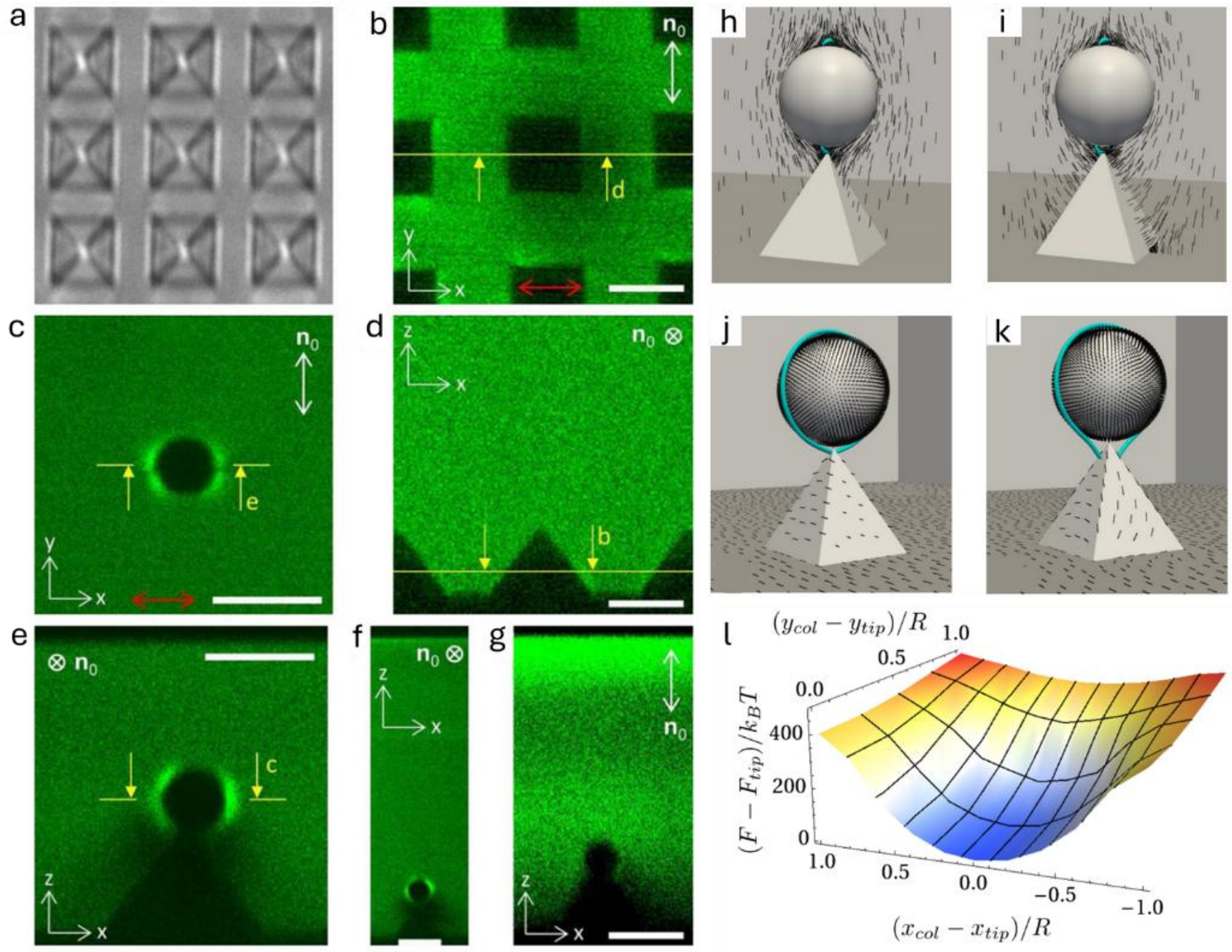

**Figure 21. Template-assisted colloidal assembly in LC.** a-g, Experimental textures of particles elastically trapped at the tip of the pyramids patterned at the glass substrate: unpolarized bright field (a) and a nonlinear luminescence image in-plane $(x, y)$ (b) and cross sectional $(z, x)$ (d) textures of a substrate patterned with a pyramid array in a planar liquid crystal cell. c, A silica particle with homeotropic anchoring and a Saturn ring defect elastically trapped on the tip of a pyramid with planar anchoring in the thin (e) and thick (f) nematic cell. g, A melamine resin particle with tangential degenerate anchoring elastically trapped on the tip of a pyramid with homeotropic anchoring. White double arrow and crossed circle show a far-field director $\mathbf{n}_0$, respectively, in plane and out of plane of the image. Dark red double arrow shows the direction of the polarization of excitation beam. Scale bar is 5 $\mu$m. h-k, Computer-modelling of colloidal particles trapped at the tips of pyramid protrusions with the centers of the particles being aligned with the pyramid tips along $z$ – direction. h and i, Colloidal particles impose tangential degenerate anchoring, while homeotropic anchoring is set at the patterned surface. j and k, Homeotropic anchoring at the particle surface and tangential degenerate one at the patterned substrate. Regimes of a vanishing (h, j) and strong (i, k) anchoring strengths at the patterned surfaces are presented. Numerically calculated $\mathbf{n}(\mathbf{r})$ is shown by short black lines. Isosurfaces of reduced scalar order parameter enclose cores of topological defect lines and are shown by blue tubes. L, Landau–de Gennes free energy as a function of the lateral position $(x_{col}, y_{col})$ of the particle with radius $R$, and with tangential degenerate anchoring at a pyramid-decorated substrate with homeotropic anchoring. The $z$ – coordinate of the particle $z_{col} = z_{tip} + 1.1R$, where $z_{tip}$ is the $z$ – coordinate of the pyramid tip; and $F_{tip} \equiv F(0,0)$. Reproduced with permission from [10].

### 6.5. Topological textures as particle-like building blocks

Skyrmions, torons, and hopfions are often discussed primarily in terms of topological classification, but their importance in soft matter is also tied to their behavior as localized mesoscale excitations. They can be created individually, positioned, driven, arranged into arrays, and made to interact through the distortions they induce in the surrounding orientational field [1]. Torons, in particular, have been shown to form chains and periodic structures and to respond collectively to applied electric fields [40,41]. More generally, such localized chiral textures can behave as field-based mesoscale objects rather than as static configurations to be classified only by their topological invariants.

This broadens the concept of self-assembly developed in the previous section. The assembling unit no longer needs to be a fabricated colloidal particle; it can also be a localized texture written directly into the medium. In this sense, liquid crystals provide a route from colloidal assembly to solitonic assembly. Whether the relevant entity is a solid inclusion or a localized field configuration, the same basic mechanism applies: geometry, topology, anchoring, and chirality determine the structure of the underlying field, and that field mediates the interaction. Similar ideas are now emerging in chiral magnetic systems, where hopfions and related three-dimensional solitons can likewise behave as localized interacting objects [33]. Rather than representing a sharp distinction between soft and hard condensed matter, these developments point to a broader unifying principle: ordered media can support particle-like topological textures whose interactions and collective organization are governed by the geometry and topology of the underlying field.

The main conclusion is that chirality is not merely a source of additional defect variety. It provides a route to localized topological entities—some singular, some nonsingular—that derive their stability from confinement and their interactions from the surrounding elastic medium. This leads naturally to the final theme: defects, colloids, and solitonic textures can all serve as programmable mesoscale elements whose interactions are tuned by anchoring, chirality, confinement, and external fields.

## 7. From topology to self-assembly and materials design

The preceding sections have shown how curvature, topology, anchoring, and chirality shape director fields and the defects or solitons they contain. The next step is to understand how these field structures mediate effective interactions and, through them, generate assembled material architectures. This is the central materials message of the liquid-crystal-colloid perspective developed here: geometry determines distortion symmetry, distortion symmetry determines interaction symmetry, and the resulting interactions guide self-assembly. Topological liquid-crystal systems therefore offer a route to composite materials whose organization is encoded not only in chemistry, but also in the topology of inclusions and of the surrounding field [1,2].

This viewpoint shifts the emphasis from passive geometric influence to active geometric design. In liquid crystals, curvature can do more than modify pre-existing structures: it can help generate the very building blocks of assembly, including defects, elastic multipoles, and localized topological textures. The resulting architectures are therefore not merely shaped by geometry; they emerge from the way geometry organizes the orientational field. The central lesson is that geometry acts through the director field to create interaction units, whose collective behaviour then gives rise to larger self-assembled soft-matter structures [1,2].

### 7.1. Elastic interactions and design rules for mesoscale self-assembly

In a nematic host, every inclusion or localized texture generates a long-ranged elastic distortion field. Even ordinary spherical colloids can give rise to dipolar, quadrupolar, or higher-order interaction symmetries, depending on anchoring and defect structure. When the inclusion has nontrivial topology or deliberately designed geometry, the surrounding elastic field becomes correspondingly richer. Shape-controlled and boundary-condition-controlled inclusions may therefore be viewed as programmable elastic multipoles whose pair interactions select preferred approach angles, bonding motifs, and assembled orientations. This idea was established clearly in work on shape-controlled colloidal interactions and was later extended to particles of greater geometric and topological complexity [2,17].

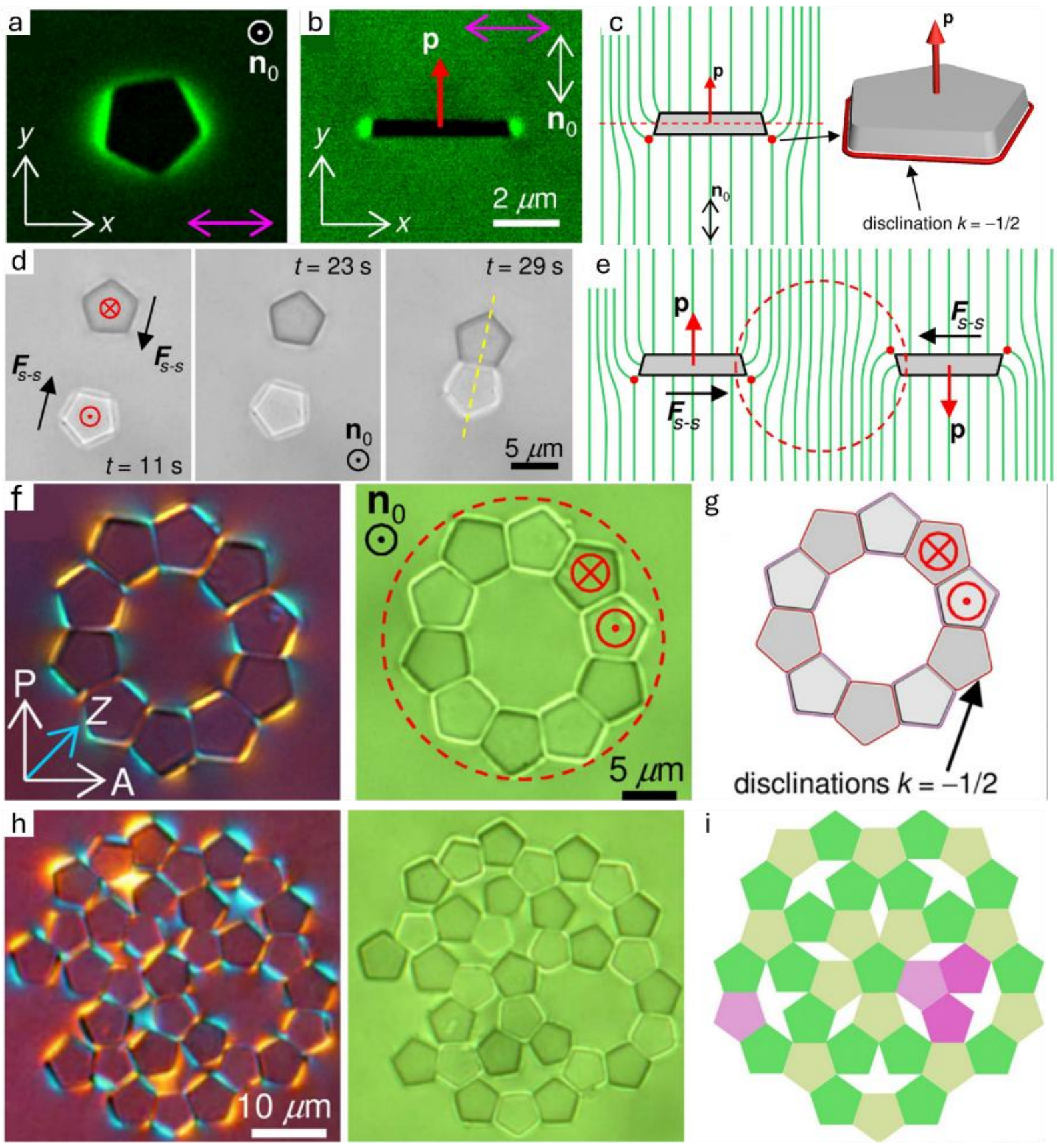

**Figure 22. Geometry-guided self-tiling in nematic colloids.** a and b, A nonlinear luminescence image of a pentagonal truncated pyramid with induced $\mathbf{n}(\mathbf{r})$ distortions; induced elastic dipole moment $\mathbf{p}$ is shown by the red arrow in b. A magenta double arrow shows linear polarization of the luminescence excitation light. c, Schematic of $\mathbf{n}(\mathbf{r})$ (green lines) around the particle, an accompanying disclination loop is shown by red circles and a red solid line in the right panel. d and e, Side-to-side elastic pair interactions of the particle in a homeotropic nematic cell. A sequence of optical bright field micrographs showing the attractive pair interaction between antiparallel elastic dipoles (d); and the schematics of $\mathbf{n}(\mathbf{r})$ around interacting particles (e). f-i, Examples of assemblies of convex pentagonal particles in a homeotropic nematic cell. f, Optical images and g a schematic of a ringlike assembly; $P$, $A$, and $Z$ in f mark, respectively, the polarizer, analyzer and a slow axis of a 530-nm phase retardation plate. h, Micrographs of a Penrose tiling fragment, and the corresponding schematic (i) where magenta pentagons represent particles missing in the experimental assembly. Reproduced with permission from [42].

The relevance of topology for self-assembly lies in the fact that the same constraints that organize defect structure also organize interaction anisotropy. A handlebody under homeotropic anchoring, a toroidal colloid decorated by boojums, and a faceted particle with edge-pinned defects do not merely present different geometries to the eye; they imprint different elastic signatures on the surrounding nematic and thereby promote different assembly pathways. Topological classification is therefore not merely a language of description. It is a mechanism by which shape and topology are transmitted, through the director field, into effective interaction rules [1,2,5,6].

A convenient way to think about this is to regard the director field as an interaction mediator. A particle does not interact with its neighbours only through contact or short-range chemistry. Instead, it imposes a deformation on the host medium, and that deformation is sensed by other inclusions over distances much larger than the particle size. Because the distortion pattern depends strongly on topology, anchoring, and geometry, the interaction becomes programmable in ways that would be difficult to realize in isotropic fluids. This is one of the major advantages of liquid-crystal colloids as a route toward designed mesoscale organization.

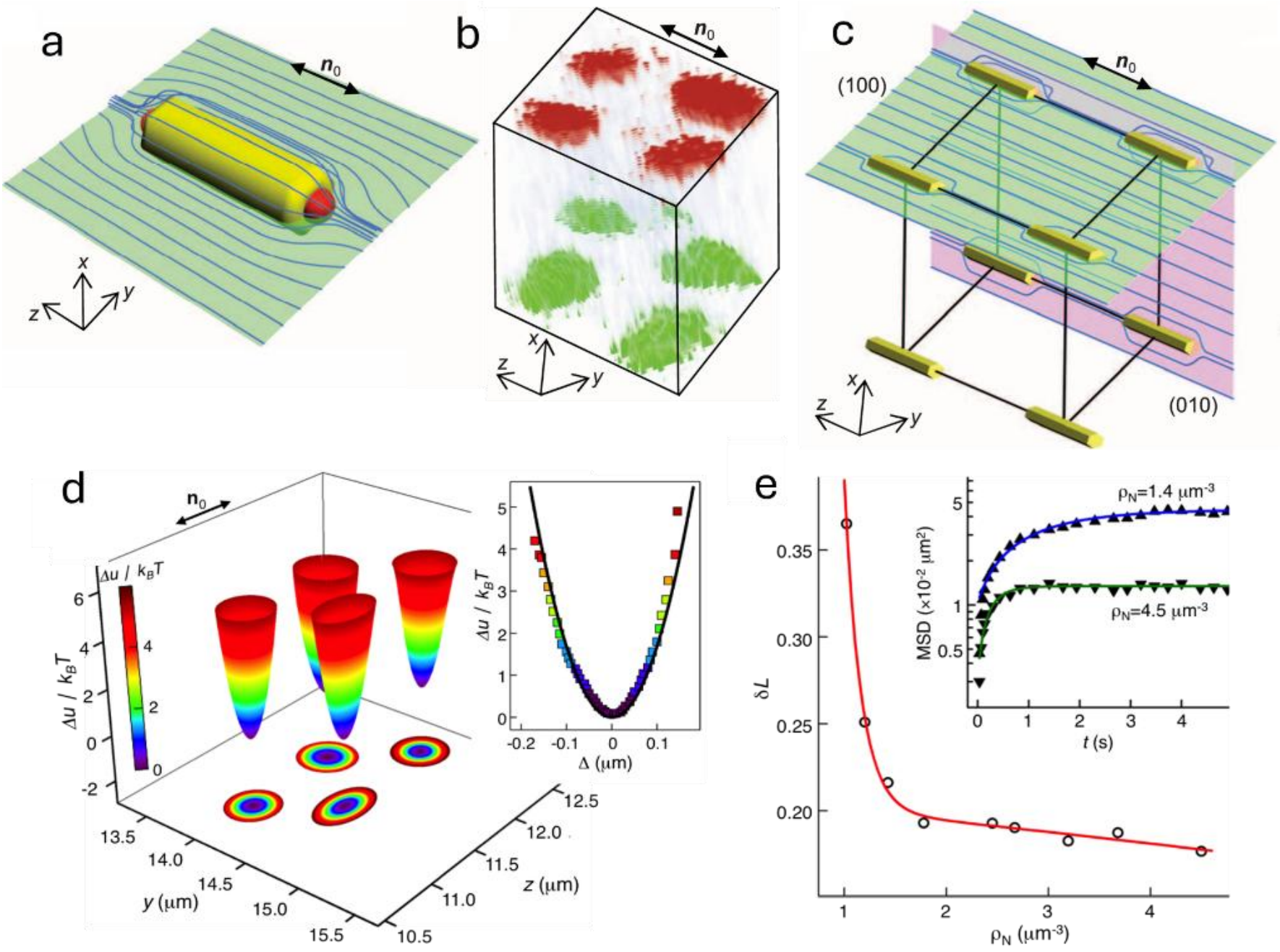

**Figure 23. Triclinic colloidal crystals.** a, Schematic representation of $\mathbf{n}(\mathbf{r})$ (blue lines) induced by a single nanorod (yellow cylinder), with the red hemispheres at the poles depicting two particle-induced boojums. This $\mathbf{n}(\mathbf{r})$ exhibits quadrupolar symmetry at far-field and is axially symmetric with respect to the nanorod axis (parallel to $\mathbf{n}_0$) and has mirror symmetry planes both parallel and orthogonal to $\mathbf{n}_0$. b, 3D micrograph showing a primitive cell of a triclinic colloidal crystal, which was reconstructed based on confocal imaging. It shows spatial arrangements of particles as they explore potential energy landscape near their minimum-energy triclinic lattice sites. c, A schematic drawing (not to scale) of the primitive cell of a triclinic crystal showing director distortions (blue lines) induced by nanorods. d, The potential energy $\Delta u$ landscape corresponding to sites in the (100) crystallographic plane of the triclinic crystal. Inset shows $\Delta u$ as a function of the deviation $\Delta$ of a particle from a local minimum. e, Lindemann parameter $\delta L$ as a function of the number density $\rho_N$ of the nanorods, characterizing the crystallization-melting transition. Inset shows the MSD of particles versus time at number densities 1.4 $\mu m^{-3}$ (▲) and 4.5 $\mu m^{-3}$ (▼). Reproduced with permission from [44].

### 7.2. Geometry-guided and template-assisted assembly

One of the clearest demonstrations that patterned confinement can be used as an assembly tool is template-assisted assembly of nematic colloids (figure 21). In this approach, a sculpted substrate does not trap the particle primarily by geometric fit. Instead, convex pyramid-like protrusions generate localized director distortions that act as elastic trapping sites for colloids with complementary anchoring conditions [10]. Experiments and Landau–de Gennes modeling show that particles are preferentially localized near pyramid tips, where the elastic distortions and defect structures associated with the particle and substrate can be matched (figures 21e–k) [10]. The corresponding free-energy landscape has a well-defined trapping minimum (figure 21l), producing a lock-and-key-like mechanism governed by elastic and topological compatibility rather than by steric shape alone.

The broader significance is that geometric frustration is converted into a constructive design principle. A patterned surface can prescribe regions of strong director deformation, while a colloidal inclusion supplies a complementary distortion field. This allows one to program not only pair interactions between freely suspended particles, but also the spatial placement of particles within a structured environment. Importantly, this strategy targets the defect and elastic field rather than the particle body itself, so trapping can be strong while remaining potentially reversible when the liquid crystal is melted or reconfigured [10].

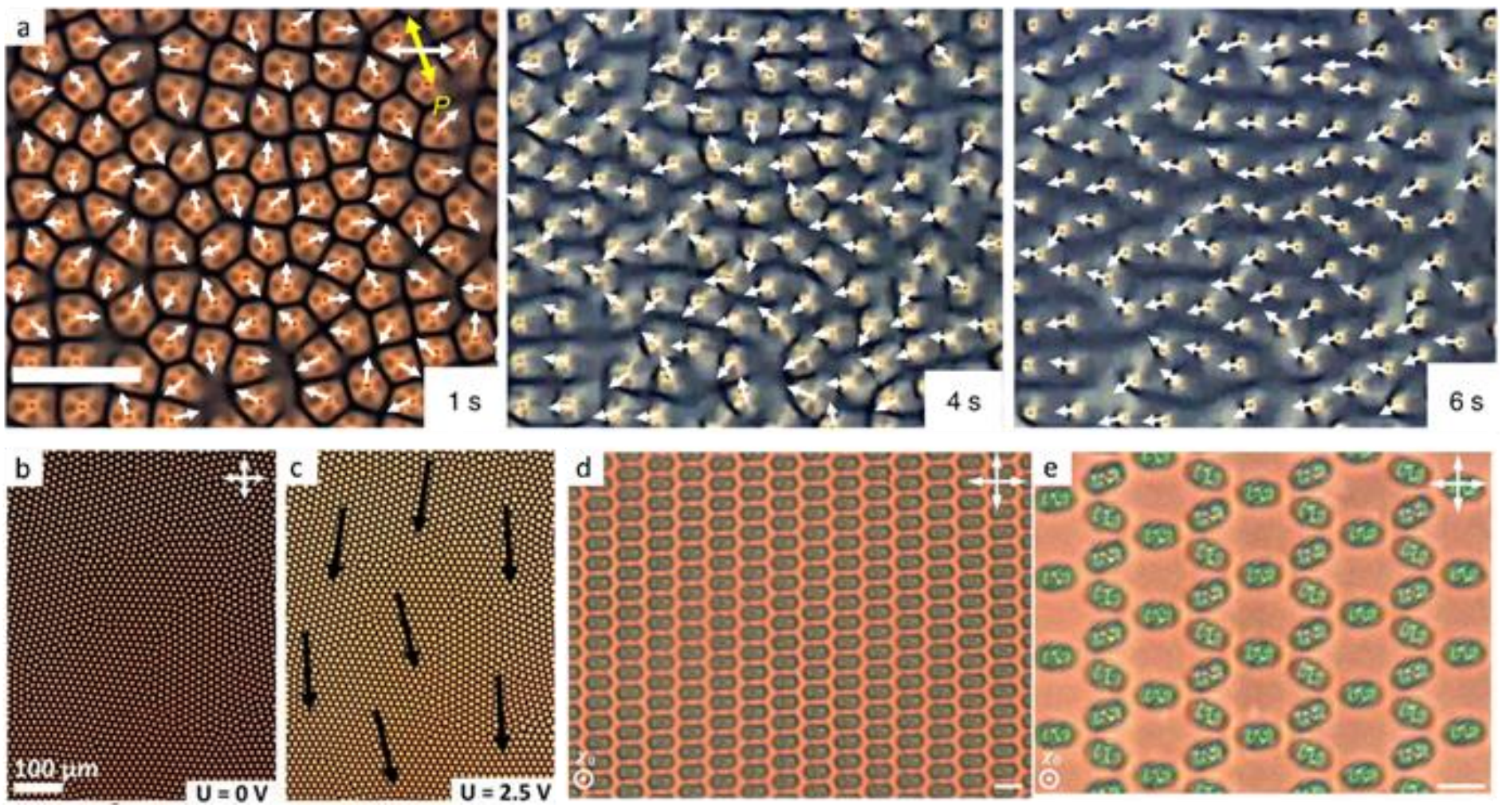

**Figure 24. Emergent collective behaviour of topological solitons in LC.** a, Polarising micrograph showing a temporal evolution of skyrmion velocity vector orientations in a high number density region. The scale bar is $100\ \mu m$; adapted with permission from [40]. b and c, Close-packed crystallites of torons are shown in polarising micrographs at zero voltage $U = 0$ (b) and at $U = 2.5\ V$ (c). Grain boundaries separating different crystallites are clearly visible in b. The black arrows in c denote the directions of the collective motion of each crystallite; adapted with permission from [41]. Closed (d) and open (e) rhombic heliknoton lattices obtained at applied voltage $U = 1.9\ V$ and $U = 1.7\ V$, respectively; the scale bars correspond to $10\ \mu m$. Reproduced with permission from [31].

This templating route also connects naturally to fabrication. Arrays of pyramidal protrusions can define preferred nucleation sites for colloidal chains or crystals, thereby setting the orientation and symmetry of larger assemblies. More generally, a colloidal particle with a given anchoring condition may be viewed as a mobile elastic-topological element, while the patterned substrate acts as a fixed elastic landscape. Matching between the two provides a soft-matter analog of lock-and-key recognition, with the "key" defined by director distortions, defect placement, and anchoring compatibility rather than by molecular complementarity alone [10].

A closely related route is geometry-guided self-tiling, in which particle shape and the accompanying defect structure jointly determine the assembly pathway. Truncated polygonal pyramids with homeotropic anchoring form elastic dipoles in the nematic host, with a $-1/2$ disclination loop localized around the larger-base edge (figures 22a–e). At large separations these particles interact in an electrostatic-dipole-like manner, whereas at short range their polygonal shape and defect-loop geometry select side-to-side or base-to-base binding modes. Experiments show that antiparallel elastic dipoles attract side-to-side, while parallel dipoles repel in the same geometry; in planar cells, base-to-base attraction provides an additional assembly channel [42]. These interactions allow pentagonal and rhombic truncated pyramids to self-tile into chains, closed rings, vertex stars, crystalline fragments, and quasicrystal-like Penrose tiling motifs; figures 22f–i show some of the possible assemblies of pentagonal particles [42].

The broader implication is that colloidal self-assembly can be programmed by designing the particle and its elastic environment as a coupled system. The resulting architecture is determined neither by particle shape alone nor solely by the far-field elastic multipole. Instead, it reflects the combined influence of shape, anchoring, defect placement, and long-range director distortion. Geometry-guided self-tiling therefore links ordinary shape-directed packing with liquid-crystal-mediated assembly: the particle provides the geometric motif, while the nematic host converts that motif into anisotropic elastic interactions. A similar idea appears in theoretical work on key-lock colloids, where shape complementarity and elastic compatibility are treated together as ingredients of selective binding [43]. Together, these studies suggest a broader inverse-design strategy: specify the desired assembly first, and then choose the particle geometry, anchoring, curvature, and topology needed to realize it.

### 7.3. From pair interactions to colloidal crystals

The design logic becomes especially compelling when it leads to bulk periodic order. Nematic colloids have long been known to form chains and low-dimensional aggregates, but a major advance was the realization that competing long-range interactions can stabilize genuine three-dimensional colloidal crystals with very low crystallographic symmetry [2]. Triclinic nematic colloidal crystals provide a clear example. In these systems, colloidal nanorods with boojums generate quadrupolar elastic distortions in the surrounding director field (figure 23a). Their assembly is governed by the competition between anisotropic elasticity-mediated interactions and weakly screened electrostatic repulsion, producing sparse three-dimensional lattices with micrometer-scale unit cells (figure 23b and c) [44].

This result shows how local director distortions can scale up into mesoscale matter. The symmetry of the particle-induced distortion field, together with the long-range repulsive interaction, defines an energy landscape with well-separated minima at the triclinic lattice sites (figure 23d). Thermal fluctuations around these minima, together with the crystallization–melting transition quantified by the Lindemann parameter and mean-square displacement, can be analysed using concepts closely related to those used for atomic crystals (figure 23e). The important difference is that here the “bonding” is not set by nearest-neighbour packing alone, but by the combined elastic and electrostatic interaction landscape of the nematic host.

A striking feature of this system is that the crystal is both sparse and symmetry-selective. The lattice spacing can greatly exceed the particle size, while the crystallographic axes follow the background director, which can itself be controlled using methods familiar from liquid-crystal device technology. This suggests a route to colloidal solids whose mechanical, optical, and transport-like properties are governed not simply by volume fraction, but by a deliberately structured interaction network. Triclinic nematic colloidal crystals therefore extend the same design principle from individual inclusions and pair interactions to bulk soft materials with pre-engineered architecture.

### 7.4. Solitons as self-assembling meta-particles

The route from topology to assembly is not limited to solid inclusions. As discussed in Section 6, chiral localized textures such as skyrmions, torons, and hopfions can themselves behave as particle-like mesoscale objects. They can be generated, moved, assembled, and reconfigured by optical or electric fields, while their interactions are mediated by elastic distortions of the surrounding director field [1,30,38-41]. In this case the “particle” is not a fabricated solid body, but a localized metaatom-like field configuration whose stability and interactions arise from confinement, chirality, and topology.

This becomes especially clear in multi-soliton systems (figure 24). Skyrmions can exhibit collective velocity alignment and schooling-like dynamics at high number density (figure 24a) [40]. Dense toron arrays can form close-packed crystallites whose collective motion is driven by applied voltage; under these nonequilibrium conditions the crystallites deform anisotropically, grain boundaries evolve, and orientational order can increase during motion (figures 24b and c) [41]. Heliknotons provide a further example: these three-dimensional solitons combine a nonsingular Hopf-like structure in the director field $\mathbf{n}(\mathbf{r})$ with knotted structure in the helical-axis field $\boldsymbol{\chi}(\mathbf{r})$, and can assemble into open or closed rhombic lattices depending on applied voltage (figures 24d and e), as well as the lowest symmetry 3D triclinic crystals [31].

These examples blur the usual distinction between a material particle and a field texture. In many condensed-matter systems that distinction is sharp: particles are discrete objects, while field configurations describe the medium around them. In liquid crystals, by contrast, localized textures can acquire many of the roles usually assigned to particles. A "particle" may be a fabricated inclusion, a droplet, a defect complex, or a largely nonsingular soliton. What unifies these diverse cases is the underlying elastic and topological structure of the orientational field, which defines both the identity of the localized object and the nature of its interactions.

Accordingly, controlled self-assembly extends beyond the organization of externally fabricated colloids. The assembly units can also be written directly into the medium as localized topological textures. Whether the object is a solid inclusion, a defect complex, or a solitonic field configuration, the same principle applies: topology, confinement, and chirality create a localized entity, while the surrounding orientational field determines how it interacts with others. This shifts self-assembly from particles embedded in a medium to include particle-like states of the medium itself [1,38,39].

### 7.5. Toward functional composite materials

The motivation for these studies extends beyond the classification of complex topological configurations; it seeks to utilize them as practical routes to material function. Liquid-crystal colloids have been framed explicitly as platforms for mesostructured composite materials, photonics, and topological design [2]. By organizing particles, defects, and solitons in controlled ways, these systems serve as scaffolds for hierarchical optical anisotropy, mechanically responsive architectures, and reconfigurable diffraction elements, as well as templates for the spatial organization of nanoparticles. The significance of topological colloids and templates lies in their capacity to convert abstract geometric constraints into experimentally addressable material properties [2,10].

A notably attractive feature of this approach is reconfigurability. Because the host medium is a soft, field-responsive fluid, the resulting assemblies can often be tuned or transformed by temperature, electric fields, magnetic fields, or optical manipulation. This distinguishes the soft-matter approach from many solid-state realizations of geometric effects, where the architecture is fixed during fabrication. In liquid crystals, while a boundary or inclusion may impose a specific geometry, the resulting director field configurations remain switchable. Topological self-assembly is therefore not only a route to static architectures, but also to adaptive and responsive materials [1,2,38].

One may further regard these systems as versatile testbeds for inverse materials design. Rather than starting with a given particle shape and observing the resulting assembly, researchers can begin with a target texture or architecture and determine the specific combination of topology, anchoring, and curvature required to realize it. While this program remains in its developmental stages, the studies reviewed here suggest that liquid-crystal colloids and confined nematics provide an ideal setting for implementing such strategies in a practical and scalable manner [44].

### 7.6. Concluding perspective

Taken together, the studies surveyed here suggest a hierarchical framework for designing soft matter under geometric control. At the foundational level, curvature and topology define the admissible director configurations and the associated defect charge. At the intermediate level, these configurations determine the effective interactions between colloidal inclusions or localized textures. At the emergent level, those interactions drive self-assembly into complex architectures, including low-symmetry clusters, superlattices, and templated chains. This hierarchy makes liquid-crystal colloids and confined nematics especially valuable model systems: the full progression from geometry and topology to physical interactions and organized matter can be followed, controlled, and imaged directly in real space [1,2].

The overarching conclusion is that curvature in soft matter is not a passive constraint, but a design principle. When combined with anchoring, topology, and chirality, it enables defects, solitons, and elastic interaction patterns to be written deliberately into liquid-crystalline media. These structures can then serve as programmable units for self-assembly, whether they are tied to solid colloids, droplets, defect networks, or localized field textures. The examples discussed here therefore illustrate a general route from geometry to function: curvature shapes the orientational field, the field defines interactions, and those interactions generate organized, reconfigurable soft materials.

## 8. Outlook

The developments surveyed here show that curvature, topology, anchoring, and chirality are becoming elements of a coherent design language for soft matter. At the same time, many of the strongest ideas in the field remain closer to paradigms than to fully generalized design rules. The next advances will likely come not simply from adding new examples of textures or particle shapes, but from learning how to move predictively between geometry, topology, and material function. Several directions appear especially promising.

A first open problem is the predictive inverse design of curved and topological colloids. Much of the existing literature proceeds in the forward direction: one chooses a particle shape, genus, anchoring condition, or confinement geometry, and then determines the resulting director field, defect set, and interaction symmetry. The inverse problem is considerably more difficult and, for materials design, more significant. Given a target interaction symmetry, a desired binding motif, or a sought-after mesoscale architecture, what particle topology and local geometry must be chosen to realize it? Answering that question will require more than formal topological classification. It will require quantitative maps from shape, anchoring, and elastic constants to defect morphology, metastability, and pair interaction energy. The broader opportunity is clear: if such maps become reliable, topological colloids could be designed with the same rigor as patchy particles or metamaterial unit cells, with the director field itself acting as the programmable interaction medium [44,45].

A second open problem is how curvature couples to chirality and activity. Chirality has appeared above mainly as a mechanism for stabilizing localized topological textures, often under confinement. Activity adds a further nonequilibrium ingredient, shifting the problem from static state selection to sustained motion, flux, and defect production. Although curved active nematics are known to display geometry-sensitive defect dynamics, the regime in which activity, chirality, and nontrivial topology act together remains largely unexplored. This combination could generate states unavailable in equilibrium systems, such as self-propelled or circulating localized textures, defect pumping around handles and boundaries, or curvature-biased active flows that select one handedness over another. The central question is whether curvature in such systems merely selects metastable structures, or whether it can control persistent topological dynamics [46].

A third direction concerns how topological defects and textures respond to hydrodynamic driving. Most of the systems discussed here are studied either in static equilibrium or in gently driven conditions where the defect structure remains close to a quasistatic minimum. Yet many soft-matter applications involve shear, Poiseuille flow, interfacial advection, or dynamically reconfigurable environments. This raises a set of fundamental questions. Which topological structures survive under flow, and which are transformed, reconnected, or annihilated? How robust are handlebody-induced defect networks, skyrmions, torons, or linked disclinations when the medium is continuously advected? Can flow itself become a design parameter for switching between singular and nonsingular realizations of the same topological requirement? These questions are essential because flow does not simply perturb a texture; it couples to the orientational field and to defect mobility, reshaping the kinetic pathways by which topology is expressed. A mature theory of curvature-guided soft matter must therefore include topological response under hydrodynamic driving [47].

A fourth promising direction is the use of defects and localized textures as templates for functional nanoparticles. The examples discussed above show that these structures are not only signatures of geometry; they can also act as mesoscale organizing centers. The next step is to exploit them more deliberately as scaffolds for positioning functional components. Nematic defects, disclination loops, boojums, and chiral localized textures provide regions of modified order and elastic stress that can trap, guide, or concentrate nanoparticles with optical, plasmonic, catalytic, or magnetic functionality. This suggests a route to hybrid materials in which curvature and topology first define the defect architecture, which in turn dictates the placement of functional inclusions. In this scenario, defects cease to be merely unavoidable consequences of topological constraint and become addressable templates for materials integration [48].

A fifth, and increasingly timely, direction is machine-learning-assisted inference of topology from optical textures. One of the great advantages of liquid crystals is that their textures can be imaged directly, but interpreting these signals remains difficult, especially when the structures are three-dimensional, metastable, or only partially visible. Recent work indicates that machine-learning methods are beginning to play a transformative role in the classification of phases and textures, the analysis of defects, and the inference of material properties [49]. Automated recognition of topological defects in complex nematic images is now becoming feasible [50]. In parallel, convolutional neural networks have been used to infer physically relevant parameters directly from the optical textures of liquid-crystal skyrmions, demonstrating that the microscopy signatures of topological structures can be turned into quantitative inputs for data-driven analysis [51]. This would accelerate both experiment and design by making topology inference faster, less subjective, and more scalable across large data sets.

Viewed together, these five directions point toward a common future. The field is moving from demonstration to control, from isolated examples to systematic design, and from static classification to dynamically programmable soft matter. If this transition succeeds, liquid-crystal colloids and confined nematics will be valued not only as visually compelling realizations of topology, but as platforms in which curvature, topology, chirality, flow, and data-driven inference are combined into a practical framework for designing responsive and functional mesostructured matter.

## Acknowledgements

I.I.S. would like to thank his students and postdocs at the University of Colorado Boulder who have been working on research topics related to this review over the years, including P.J.Ackerman, O. Iadlovska, H.Mundoor, C. Meng, A.J. Hess, N. Golden, H.R.O. Sohn, J.-S.B. Tai, J.-S. Wu, B. Senyuk, J. Brewer, Z. Chen, J.S. Evans, M.G. Campbell, J. Bourgeois, T. Boyle, A. Pattis, C.D.Liu, Q.Liu, A.Martinez, H.C.Mireles, Y.Wang, A.J. Funk, M.J. Laviada, I. Klevets, C. Lapointe, J. Giller, S. Park, O. Puls, D. Glugla, M.C.M. Varney, Q. Zhang, R. Voinescu, C. Twombly, G.H. Sheetah, A.J. Seracuse, M.B. Pandey, R.P.Trivedi, N.P. Garido, R. Visvanathan, O. Trushkevych. I.I.S. also acknowledges the financial support of research on Chiral LC colloids (the U.S. Department of Energy, Office of Basic Energy Sciences, Division of Materials Sciences and Engineering, under Award ER46921, contract DE-SC0019293 with the University of Colorado at Boulder) and topological solitons (the U.S. National Science Foundation grant DMR-1810513), examples of which have been prominently highlighted within this review and used in many figures. MT acknowledges financial support from the Portuguese Foundation for Science and Technology (FCT) under the contracts no. UID/00618/2025 (DOI: 10.54499/UID/00618/2025); UID/PRR2/00618/2025 (DOI: 10.54499/UID/PRR2/00618/2025); and UID/PRR/00618/2025 (DOI: 10.54499/UID/PRR/00618/2025).